\documentstyle[11pt]{article}

\sf
\begin{document}

\input psfig

\hbox{}
\vskip .50truecm

\centerline{\Large \bf A linear thermohaline oscillator driven by}
\centerline{\Large \bf  stochastic atmospheric forcing}

\vskip 1.0truecm

\centerline{Stephen M. Griffies{\dag} and Eli Tziperman{\ddag}}

\vskip 1.0truecm

\centerline{{\dag} Princeton University}
\centerline{Atmospheric and Oceanic Sciences Program}
\centerline{Sayre Hall, Forrestal Campus}
\centerline{Princeton NJ 08544-0710. USA}
\centerline{email: smg@gfdl.gov}

\vskip 1.0truecm

\centerline{{\ddag} Princeton University}
\centerline{Atmospheric and Oceanic Sciences Program}
\centerline{Sayre Hall, Forrestal Campus}
\centerline{Princeton NJ 08544-0710.  USA}
\centerline{and}
\centerline{Department of Environmental Sciences}
\centerline{Weizmann Institute of Science}
\centerline{Rehovot, 76100.  ISRAEL}
\centerline{email: citziper@weizmann.weizmann.ac.il}

\vskip 1.0truecm

\centerline{Revised submission to {\em Journal of Climate}}
\centerline{Los Alamos e-print ao-sci/9502002 }

\vfill\eject


\centerline{\large\bf Abstract}

The interdecadal variability of a stochastically forced four-box model
of the oceanic meridional thermohaline circulation (THC) is described
and compared to the THC variability in the coupled ocean-atmosphere
GCM of Delworth, Manabe, and Stouffer (1993).  The box model is placed
in a linearly stable thermally dominant mean state under mixed
boundary conditions.  A linear stability analysis of this state
reveals one damped oscillatory THC mode in addition to purely damped
modes.  The variability of the model under a moderate amount of
stochastic forcing, meant to emulate the random variability of the
atmosphere affecting the coupled model's interdecadal THC variability,
is studied.  A linear interpretation, in which the damped oscillatory
mode is of primary importance, is sufficient for understanding the
mechanism accounting for the stochastically forced variability.
Direct comparison of the variability in the box model and coupled GCM
reveals common qualitative aspects.  Such a comparison supports,
although does not verify, the hypothesis that the coupled model's THC
variability can be interpreted as the result of atmospheric weather
exciting a linear damped oscillatory THC mode.

\vfill\eject

\section{Introduction}
\label{section:introduction}

The need to differentiate anthropogenic climate change from natural
climate variability has generated intense research efforts directed at
understanding and modeling climate variability.  In particular,
attention has focused on phenomena occurring on interdecadal or longer
time scales.  As it is believed that the ocean is a major climate
component determining variability at these time scales, several
ocean-only models as well as coupled ocean-atmosphere models of
varying sophistication have been employed to study such variability
and to identify its fundamental modes.

The meridional heat transport in the Atlantic basin is strongly
dependent on the thermohaline circulation (THC).  Hence, variability
in the Atlantic's THC is potentially connected with climate
variability in and around the Atlantic.  Much of the evidence for this
connection, both from models and historical data, is discussed in the
review by Weaver and Hughes (1992).  Additional reviews are provided
by Willebrand (1993) and Marotzke (1994).

Modeling efforts, from box-models (e.g., Stommel 1961) to fully
coupled ocean-atmosphere GCMs (e.g. Delworth et al. 1993; hereafter
D93), have indicated a suite of variability in the THC.  The time
scales of the model variability range from decadal and century (e.g.,
Weaver et al.\ 1991, 1993, 1994, Mikolajewicz et al.\ 1990, Weisse et
al.\ 1994, Winton et al.\ 1993, Mysak et al.\ 1993, Moore and Reason
1993, Yin and Sarachik 1993, Greatbatch and Zhang 1994, Cai et al.\
1994, and D93), associated with advective/convective mechanisms, to
millennial (e.g., Marotzke 1989, 1990, Weaver and Sarachik 1991, and
Weaver et al.\ 1993), associated with diffusive mechanisms.  The
amplitudes of the variability range from a small to rather large
percentage of the mean circulation, associated with the interdecadal
to century oscillations, to massive flushing events occurring on the
millennial time scale.  The variability studied in the ocean-only
models has been predominantly associated with self-sustained
oscillations involving non-linear mechanisms.  The models of
Mikolajewicz et al.\ 1990, Weisse et al.\ 1993, and Mysak et al.\ 1993
are exceptions in which the decadal and century variability is driven
by stochastic salinity forcing in which a linear modal interpretation
is available.

D93 have shown that the coupled ocean-atmosphere climate model
developed at NOAA's Geophysical Fluid Dynamics Laboratory exhibits
small amplitude [$\approx 5\%-10\% $ of the roughly $20$ Sv ($10^{6}
\mbox{m}^{3}/\mbox{sec}$) mean meridional circulation] interdecadal
($40-60$ years) thermohaline variability about a thermally dominant
THC in the North Atlantic sector of their model.  The basic mechanism
identified for this variability is the advection of salt and heat into
the northern ``sinking region'' creating periodically positive and
negative density anomalies closely correlated to the circulation
anomalies.  The model variability seems to be associated with
variability about a stable climate state rather than the larger
fluctuations which may have occurred in the transition to or from an
ice age (see Weaver and Hughes 1992 and references therein for a
discussion of such transitions).

The small amplitude variability seen in D93 is consistent with,
although a not proof of, a linear mechanism being responsible for the
oscillatory behavior in the coupled model.  For studying such an
interpretation, the results of simplified models can provide useful
insight.  For example, Stommel (1961) found that a linear stability
analysis of his thermohaline two--box model displayed a damped
oscillatory response in part of its parameter range.  As pointed out
by Bryan and Hansen (1992) and Ruddick and Zhang (1994), the damped
oscillatory response exists only in a saline dominant regime when the
model is forced under mixed boundary conditions. Such a regime is not
relevant for describing the present North Atlantic variability.
However, the results of Tziperman et al.\ (1994, hereafter T94)
indicated that when more degrees of freedom are considered, a damped
oscillatory response within a thermally dominant regime can result
from a box model.  The damped oscillatory regime in T94 is near their
four box model's stability transition point; i.e., the point at which
the model's THC solution becomes unstable.  They further suggested
that their realistic-geometry primitive equation ocean GCM is near its
stability transition point, possibly in the damped oscillatory regime
found in their box model.  Assuming that the variability found in D93
is due to a damped oscillatory mode, an external excitation of this
mode is needed in order to sustain the variability.  As noted in D93,
such excitations may come from the random atmospheric weather.  In
this context, Bryan and Hansen (1992) considered the stochastic
forcing of the thermally dominant state in the Stommel two-box model.

The purpose of this paper is to examine the hypothesis that the
interdecadal variability observed in the coupled model analysis of D93
can be interpreted as the excitation of a damped oscillatory
thermohaline mode by the stochastic effects of atmospheric
variability.  As a first step in testing this hypothesis, we
investigate the variability of a highly simplified mixed boundary
condition ocean-only four-box model of the oceanic meridional
circulation forced by stochastic surface heating.  A direct comparison
of the box model variability to that of the coupled model shows a
qualitative agreement between the mechanisms acting in the two models.
Such agreement supports, although clearly cannot validate, the
proposed hypothesis for the coupled model.

Some important aspects of the variability seen in the coupled model of
D93 have been captured in the recent work of Greatbatch and Zhang
(1993) using a three-dimensional planetary geostrophic ocean model
forced with a zonally symmetric fixed heat flux and a zero salinity
flux.  This work was extended by Cai et al.\ (1994) to a primitive
equation model with a zonally asymmetric fixed heat flux.  Both of
these studies identified variability of a gyre circulation in the
North Atlantic similar to that identified by D93.  However, the
simulations of Greatbatch and Zhang and Cai et al.\ do not capture the
temperature and salinity phase relations of D93 due to their use of a
fixed heat flux.  Although the two dimensional mixed boundary
condition box model studied in this paper cannot capture the
three-dimensional gyre-effect, we will argue that it does capture much
of the zonally averaged variability acting in the coupled model; in
particular, the phase relations between temperature/salinity and the
THC are quite similar to that of D93.  Both the temperature/salinity
phase relations and the gyre effect are important parts of the
variability seen in D93 and the full explanation of the coupled
model's variability probably includes both the mechanisms suggested
here and in Greatbatch and Zhang and Cai et al.  An additional
important difference between the present work and that of Greatbatch
and Zhang and Cai et al.\ is that the box model's variability is
linear and noise driven whereas that of Greatbatch and Zhang and Cai
et al.\ is self-sustained and hence nonlinear in character.  The
different perspectives taken in this study and that of Greatbatch and
Zhang and Cai et al., both of which seem to give reasonable agreement
with various aspects of the coupled model's variability, are perhaps
part of a continuing dialogue necessary to successfully understand and
simulate the ocean's THC.

We begin in Section \ref{section:4-box-model} with a detailed
description of the box model and its linear stability properties while
under mixed boundary conditions.  The essential results from this
section are summed up in Figs.\ \ref{fig:box_geometry} and
\ref{fig:combined_plots} which show the box model configuration and
its damped oscillatory response, respectively. It is the variability
about such a stable, damped oscillatory steady state that is of
interest in the following sections.  Section
\ref{section:Linear_oscillations} presents a linearized advective
mechanism describing the box model's damped oscillatory response to
small perturbations about its thermally dominant steady state.  In
Section \ref{section:Random_forcing} we analyze the response of the
box model subjected to stochastic heating on the surface boxes.  The
results from such variability are compared to the zonally averaged
results from the coupled GCM of D93.  We present further discussion in
Section \ref{section:discussion} and conclusions in Section
\ref{section:conclusions}.

\section{The box model}
\label{section:4-box-model}

Since the work of Stommel (1961), box models have proven to be a
useful tool for studying the ocean's THC because they allow for the
examination of individual processes which might be important for the
circulation's stability and variability.  Most often, and in this work
too, the boxes have no zonal extent; i.e., the model has only two
spatial dimensions which correspond to the vertical and meridional
directions.  Though highly idealized, the perspective taken here is
that box models can provide a tool for testing hypotheses which can be
directly compared against the results of more realistic models and
observations.

\subsection{Model equations}
\label{subsection:model_equations}

\begin{figure}[htbp]
\centerline{\psfig{figure=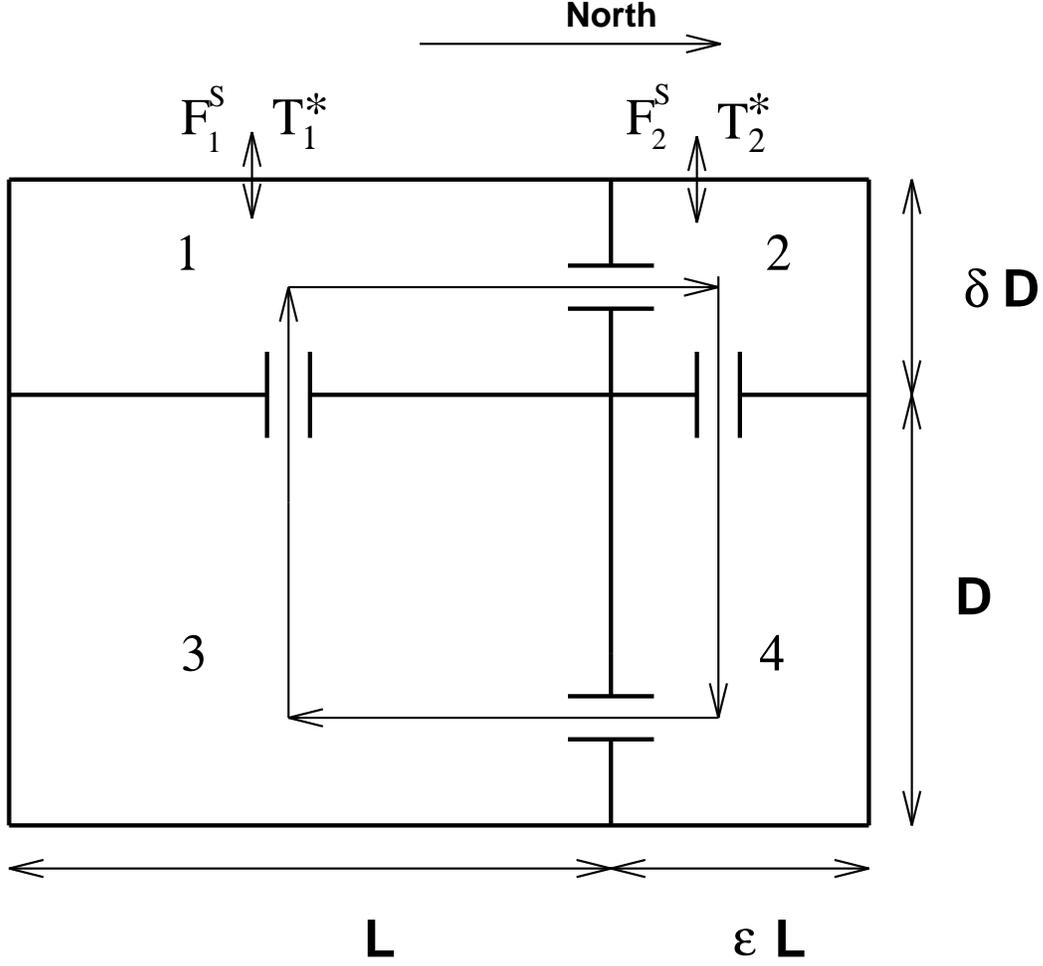,height=5.0in}}
\caption{
   \baselineskip 3ex
   Geometry of the four box model. The thermal driven
   mean circulation, with sinking in the north and rising in
   the south, is indicated.  The boxes are homogeneous.
   The parameters chosen for
   the experiments are the following: $V=8 \times 10^{16}{\mbox m}^{3},
   \epsilon=.10, \delta=.10, D=3000 {\mbox m}$ and the surface area of
   the northern box is $A = \epsilon V / D = 2.67 \times 10^{12} 
   {\mbox m}^{2}$.
   The surface box temperatures are restored to 
   $T_{1}^{*}=25^{\circ}\mbox{C}$ and $T_{2}^{*}=0^{\circ}\mbox{C}$
   with a restoring time $\gamma_{T}^{-1} = 180$ days.  
   The surface salinity forcing
   $F^{S}_{1,2}$ is restoring to the salinities $S_{1}^{*} = 36.5$psu
   and $S_{2}^{*} = 34.5$ psu 
   with a restoring time $\gamma_{S}^{-1} = 300$ days.
   Afterwards, a fixed 
   salinity flux $F^{S}_{2} = -\epsilon F^{S}_{1} < 0$ is used for the mixed
   boudary condition integrations.
   Note the four boxes all have different volumes.}  
\label{fig:box_geometry}
\end{figure}

We consider a four-box model of a thermally dominated North
Atlantic-like meridional circulation similar to those of Huang et al.\
(1992) and T94 (Fig.~\ref{fig:box_geometry}).  Relatively warm salty
boundary conditions are prescribed for the southern surface box and
cold fresh boundary conditions are prescribed for the northern surface
box.  Due to the opposing effects of salinity and temperature on the
density of sea water, salt contributes a southern sinking buoyancy
torque and temperature creates a northern sinking torque.  The model
has no local convective mixing nor any explicit diffusion; hence, the
temperature and salinity in each box are only affected by advection to
and from neighboring boxes as well as by surface fluxes in the upper
two boxes.

The steady state fields, about which variability is studied, are
computed using restoring surface conditions for both the salinity and
temperature.  Thereafter, the implied salinity flux is fixed and the
model is run with mixed boundary conditions (F. Bryan 1986).  The
temperature and salinity equations governing the box model in the
absence of stochastic forcing are
 \begin{eqnarray}
  \dot{T_{1}} &=& {U \over \delta V}(T_{3} - T_{1}) +
                \gamma_{T}(T_{1}^{*} - T_{1})
\label{eq:T1}\\
  \dot{T_{2}} &=& {U \over \epsilon \delta V}(T_{1} - T_{2}) +
                \gamma_{T}(T_{2}^{*} - T_{2})
\label{eq:T2}\\
  \dot{T_{3}} &=& {U \over V}(T_{4} - T_{3})
\label{eq:T3}\\
  \dot{T_{4}} &=& {U \over \epsilon V}(T_{2} - T_{4})
\label{eq:T4}\\
\dot{S_{1}} &=& {U \over \delta V}(S_{3} - S_{1}) + F^{S}_{1}
\label{eq:S1}\\
\dot{S_{2}} &=& {U \over \epsilon \delta V}(S_{1} - S_{2}) + F^{S}_{2}
\label{eq:S2}\\
  \dot{S_{3}} &=& {U \over V}(S_{4} - S_{3})
\label{eq:S3}\\
  \dot{S_{4}} &=& {U \over \epsilon V}(S_{2} - S_{4}).
\label{eq:S4}
\end{eqnarray}
 In these equations, the overdot indicates a time derivative, $T_{i}$
and $S_{i}$ are the box temperatures and salinities, $T_{i}^{*}$ are
the restoring temperatures for the surface boxes, $V$ is the volume of
the southern lower box $3$ (the largest box), $U$ is the circulation
volume flow rate, and $\gamma_{T}^{-1}$ is the restoring time for the
temperature. A salinity restoring condition $F^{S}_{1,2} =
\gamma_{S}(S_{1,2}^{*} - S_{1,2})$ is used for reaching a steady
state, with $\gamma_{S}^{-1}$ the restoring time and $S_{i}^{*}$ the
restoring salinities.  Afterwards, the fixed flux $F^{S}_{1,2} =
\gamma_{S}(S_{1,2}^{*} - \overline{S}_{1,2})$ is used for the mixed
boundary conditions, with the barred quantities denoting the steady
state values reached under restoring conditions.  The steady state
conditions, obtained by setting the time derivatives in equations
(\ref{eq:T1})--(\ref{eq:S4}) to zero, satisfy
$\overline{T}_{2}=\overline{T}_{3}=\overline{T}_{4} \equiv
\overline{T} <\overline{T}_{1}= T_{1}^{*} + \epsilon
(T_{2}^{*}-\overline{T})$ and
$\overline{S}_{2}=\overline{S}_{3}=\overline{S}_{4} \equiv
\overline{S} <\overline{S}_{1}=S_{1}^{*} + \epsilon
(S_{2}^{*}-\overline{S})$ for the thermally dominant flow
$\overline{U} = \delta U_{0}[(\alpha(\overline{T}_{1}-\overline{T}) -
\beta(\overline{S}_{1}-\overline{S})] > 0$ considered here.  The
dimensionless numbers $0 < \epsilon,\delta \le 1$ are geometric
factors defined in Fig.\ \ref{fig:box_geometry}.  The differencing in
equations (\ref{eq:T1})--(\ref{eq:S4}) assumes a northern sinking
thermal dominated circulation, which is that of interest in this
study.

The circulation in the box model is driven solely by horizontal
pressure gradients between the north and the south which, using the
hydrostatic approximation, correspond to density gradients.  A larger
northern density creates a positive circulation.  Therefore, the
circulation is given by
 \begin{equation}
  U = {U_0\over\rho_0}
      [ \delta(\rho_{2} - \rho_{1}) + (\rho_{4} - \rho_{3})]
      \propto (\rho_{\mbox{north}} - \rho_{\mbox{south}}),
\label{eq:circulation}
\end{equation}
 where, assuming a linear equation of state, the densities in the
boxes are $\rho_{i} = \rho_{0}[1 - \alpha(T_{i}-T_{0}) +
\beta(S_{i}-S_{0})]$ with $\rho_{0}, T_{0}$, and $S_{0}$ being the
reference density, temperature, and salinity, respectively.  The
thermal and saline expansion coefficients $\alpha$ and $\beta$ are
taken as constants throughout the system which guarantees that the
circulation (\ref{eq:circulation}) vanishes when the north--south
temperature and salinity gradients vanish.  The more realistic case in
which $\alpha$ and $\beta$ are different for the four boxes, as well
as a nonlinear equation of state, should not qualitatively change the
following results which consider only moderate variability.

The form of the advective circulation (\ref{eq:circulation}) is
commonly chosen in Stommel-type box models, such as this model, which
attempt to incorporate only advective processes.  It should be noted
that in a three dimensional circulation, east-west pressure gradients,
through geostrophy, typically are associated with meridional flow
rather than north-south gradients.  In the formulation of their
zonally averaged ocean model, Wright and Stocker (1991) provide some
heuristic motivation for using such a parametrization starting from
three dimensional equations of motion.  Additionally, the study by
Hughes and Weaver (1994) provides evidence that the meridional
circulation in their three dimensional model is directly proportional
to the north-south steric height difference, further motivating the
parametrization (\ref{eq:circulation}).  Finally, the results
exhibited in the coupled model of D93 to be described below, and the
qualitative agreement of the box model with these results, suggest
this choice for the current study.

\subsection{Box parameters}
\label{subsection:parameters}

The parameters used for the box model experiments are as follows.  In
the linear equation of state, the parameters $\alpha = 1.668 \times
10^{-4} \mbox{}^{\circ}\mbox{K}^{-1},$ $\beta = 7.61 \times 10^{-4} \;
\mbox{psu}^{-1}$, and $\rho_{0} = 1027$ kg/m$^3$ represent values for
$T_{0}=10^{\circ}\mbox{C}$ and $S_{0}=35$ psu.  The surface restoring
temperatures and salinities $T_{1}^{*} = 25^{\circ}\mbox{C}$,
$T_{2}^{*} = 0^{\circ}\mbox{C}$, $S_{1}^{*} = 36.5 \, \mbox{psu}$, and
$S_{2}^{*} = 34.5 \, \mbox{psu}$ are chosen along with the temperature
restoring time $\gamma_{T}^{-1} = 180$ days to yield reasonable steady
state conditions under restoring conditions.  The salinity restoring
coefficient $\gamma_{S} = .6\gamma_{T} (\gamma_{S}^{-1} = 300
\mbox{days}$) is chosen to give a box model response emulating that of
the coupled model of D93, as further discussed in Section
\ref{section:Random_forcing}.  The geometry of the box model roughly
reflects the region of the Atlantic from the equator northward. The
sinking in the coupled model is localized in a small region within
this northern latitude band; hence, the latitudinal width ratio
$\epsilon = .10$ (see Fig.\ \ref{fig:box_geometry}) is chosen.  The
depth ratio is $\delta = .10$ and the volume of the largest box
(southern deep box) is $V = 8 \times 10^{16} \mbox{m}^{3}$.  With this
thickness for the upper boxes (300 meters), the temperature restoring
time corresponds to that commonly taken in ocean-only GCM studies
(e.g., Tziperman and Bryan, 1993 which use 30 days for a 50 meter
upper layer). Finally, the circulation parameter $U_{0} = 8 \times
10^{4}$ Sv results in a mean circulation $\overline{U}$ on the order
of that seen in D93.

The above parameters, though chosen to roughly reflect the Atlantic
basin in the coupled model of D93, are not meant to be quantitatively
precise.  Rather, it is the purpose of the box model study to
illustrate a particular mode of variability emulating that seen in the
coupled model.  Precise quantitative correspondence is not pursued.
Along these lines, it should be noted that we have found these
oscillations to occur over quite a broad range of box model
parameters.

\subsection{Linear stability}
\label{subsection:linear_stability}

 In order to investigate the model's stability to small perturbations,
we consider the governing equations (\ref{eq:T1})--(\ref{eq:S4}) under
the mixed boundary conditions and linearized about the steady state
obtained with restoring boundary conditions;
 \begin{eqnarray}
  \dot{T_{1}'} &=& {U' \over \delta V}(\overline{T}-\overline{T}_{1}) +
{\overline{U} \over \delta V}(T_{3}' - T_{1}') - \gamma_{T}T_{1}'
\label{eq:T1prime}\\
  \dot{T_{2}'} &=& {U' \over \epsilon \delta V}\
               (\overline{T}_{1} - \overline{T}) +
               {\overline{U} \over \epsilon \delta V}
               (T_{1}' - T_{2}') - \gamma_{T}T_{2}'
\label{eq:T2prime}\\
  \dot{T_{3}'} &=& {\overline{U} \over V}(T_{4}' - T_{3}')
\label{eq:T3prime}\\
  \dot{T_{4}'} &=& {\overline{U} \over \epsilon V}(T_{2}' - T_{4}')
\label{eq:T4prime}\\
  \dot{S_{1}'} &=& {U' \over \delta V}(\overline{S}
                   - \overline{S}_{1}) +
                   {\overline{U} \over \delta V}(S_{3}'-S_{1}')
\label{eq:S1prime}\\
  \dot{S_{2}'} &=& {U' \over \epsilon \delta V}\
               (\overline{S}_{1} - \overline{S}) +
               {\overline{U} \over \epsilon \delta V}
               (S_{1}' - S_{2}')
\label{eq:S2prime}\\
  \dot{S_{3}'} &=& {\overline{U} \over V}(S_{4}' - S_{3}')
\label{eq:S3prime}\\
  \dot{S_{4}'} &=& {\overline{U} \over \epsilon V}(S_{2}' - S_{4}'),
\label{eq:S4prime}
\end{eqnarray}
 where ${T}_{i}' = (T_{i} - \overline{T}_{i}), {S}_{i}' = (S_{i} -
\overline{S}_{i})$, and ${U}' = (U - \overline{U})$ are the
temperature, salinity, and circulation anomalies, respectively.  The
set of linear equations (\ref{eq:T1prime})--(\ref{eq:S4prime}) can be
written in the matrix vector form ${\bf \vec{\dot{x}} } =
{\bf A \vec{x}}$, where the vector ${\bf \vec{x}}$ has eight components
consisting of the four temperature and four salinity anomalies.
${\bf A}$ is an $8 \times 8$ matrix dependent on the model parameters and
the underlying  steady state reached under restoring conditions
(Marotzke 1990, T94).


A stability analysis of the linear equations
(\ref{eq:T1prime})--(\ref{eq:S4prime}) was carried out in T94.  As
found there, should the salinity torque $\beta(\overline{S}_{1} -
\overline{S})$ become large enough to overcome the effects of the
temperature torque $\alpha(\overline{T}_{1} - \overline{T})$
[explicitly, if $\beta(\overline{S}_{1} - \overline{S}) > Q
\alpha(\overline{T}_{1} - \overline{T})$ with $Q \approx .4$ in the
example discussed in T94], the thermally dominant mean circulation
will be unstable to small perturbations, and such a perturbation would
result in the system making a transition to a stable saline dominant
mean state of opposite circulation direction.  Increasing the salinity
forcing in this model (e.g., using a smaller salinity restoring time
$\gamma_{S}^{-1}$) when initializing under restoring conditions
results in a larger mean salinity gradient $(\overline{S}_{1} -
\overline{S})$ and hence a larger salinity torque.  A continuum of
linear response is thus found as the salinity forcing is increased.
Namely, the stability behavior to small perturbations changes from
that of a purely exponential decay of the perturbation for small
salinity forcing, to oscillatory decaying response, to oscillatory
growing behavior, and finally, for a strong enough salinity forcing,
to small perturbations growing exponentially without any oscillations.
These stability results are further discussed in T94 where a detailed
stability analysis for the damped oscillatory regime is given.

The regime of interest in the following is that linearly stable
thermally dominant state which responds in a damped oscillatory manner
to small perturbations.  More precisely, the linearized system in the
following will contain a single damped oscillatory eigenmode (and its
complex conjugate) with the remaining eigenmodes purely damped.  The
damped oscillatory mode forms the basis for the model's variability
studied in this work.

\section{Linear thermohaline oscillations}
\label{section:Linear_oscillations}

We now examine the response of the box model, described by the
nonlinear conservation equations (\ref{eq:T1})--(\ref{eq:S4}), to an
instantaneous heat perturbation when the model is placed in a stable
damped oscillatory regime under mixed boundary conditions.  Note that
a forward Euler time stepping scheme is employed for the numerical
realizations of the box model equations (\ref{eq:T1})--(\ref{eq:S4})
with 365 time steps per model year.  There are no noticeable
differences between these results and those obtained when we used a
leapfrog scheme with intermittent Euler forward stepping to suppress
the computational mode.  As the numerical results are consistent with
the eigenvalue analysis discussed in the previous section, we are
confident that the variability reported here is that of the underlying
physical processes acting in the box model and are not adversely
affected by numerical artifacts such as truncation and numerical
dissipation.

\begin{figure}[htbp]
\centerline{\psfig{figure=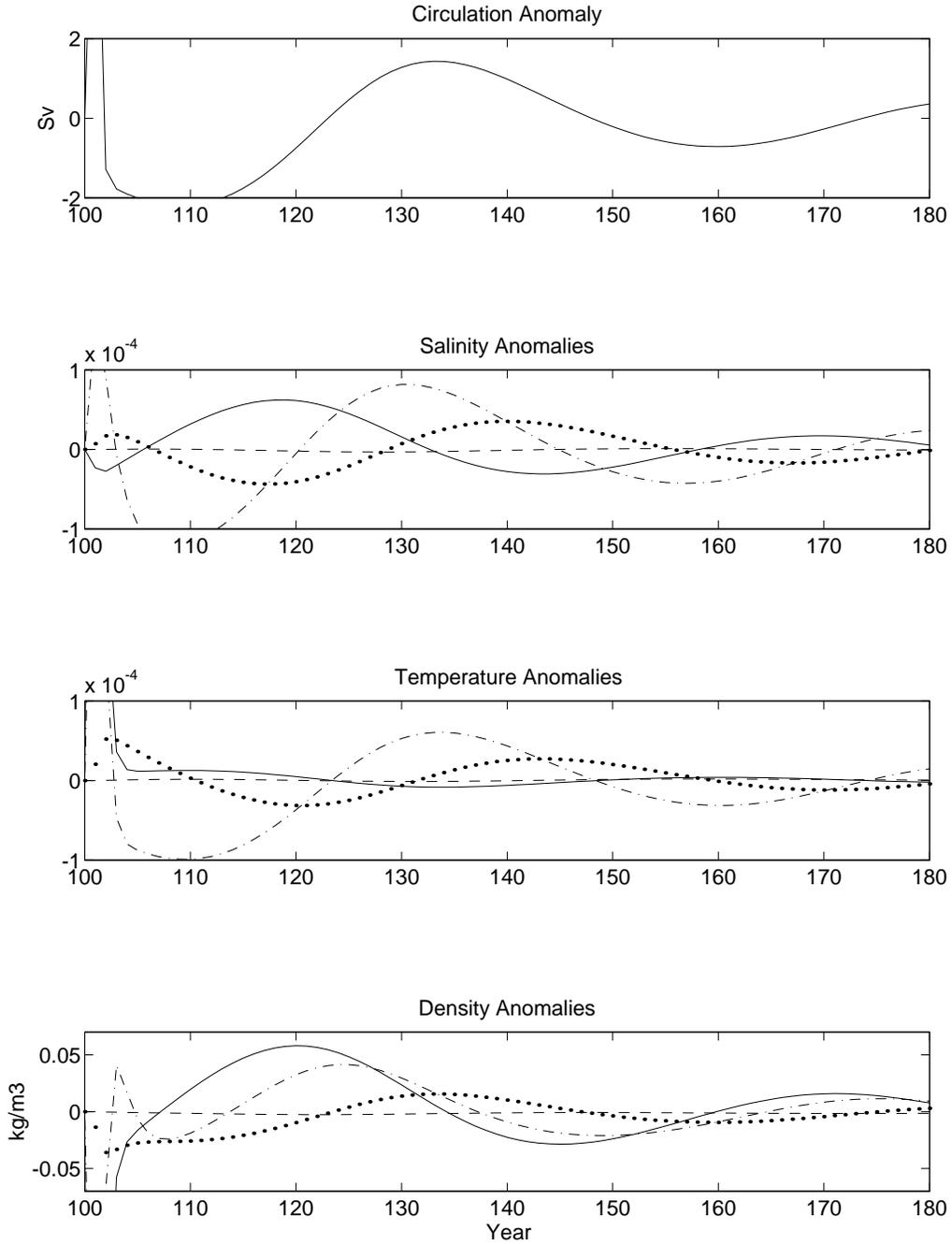,height=7.0in}}
\caption{
   \baselineskip 3ex
   Time series for the anomalous (A) circulation $U'$, 
   (B) salinity $\beta S_{i}'$, 
   (C) temperature $\alpha T_{i}'$, and (D) density $\rho_{0}(-\alpha
   T_{i}'+\beta S_{i}'$) 
   as the box model is perturbed from its initial state by the addition of 
   a cold anomaly in the north and a warm anomaly in the south.
   For the salinity, temperature, and density anomalies, 
   the solid line is for box $1$, the dot-dashed line is for
   box $2$, the double-dashed line is for box $3$, and the 
   dotted line is for box $4$.   The thermally
   dominant  mean state has a circulation of $16.9$ Sv.
   The response of the model, soon after the initial perturbation, is dominated 
   by the damped oscillatory eigenmode of the linearized system.
   Hence, the structure of this eigenmode is basically that of the plots shown
   here which were obtained from integrating the nonlinear equations
   (\protect\ref{eq:T1})--(\protect\ref{eq:S4}).   
   The period and decay times for this eigenmode 
   are $51$ years and $44$ years, respectively.} 
\label{fig:combined_plots}
\end{figure}

  Fig.~\ref{fig:combined_plots} shows the circulation, salinity,
temperature, and density anomalies for the mixed boundary condition
model as it responds to a relatively large perturbation consisting of
a cold surface anomaly in the north (box $2$) and a warm surface
anomaly in the south (box $1$).  The steady state conditions are
$\overline{T}_{1} = 24.4 \hbox{}^{\circ}\mbox{C}, \overline{T} = 6.0
\hbox{}^{\circ}\mbox{C}, \overline{S}_{1} = 36.4 \, \mbox{psu},
\overline{S} = 35.2 \, \mbox{psu}$, and $\overline{U} = 16.9$ Sv.  The
temperature and salinity anomalies shown in Fig.\
\ref{fig:combined_plots} are multiplied by their respective expansion
coefficients $\alpha$ and $\beta$ allowing for a direct comparison of
the relative contributions each field makes to the density anomalies
and hence to the circulation anomaly.

The result of the perturbation is a positive circulation anomaly.  The
phases and relative amplitudes shown in Fig.~\ref{fig:combined_plots}
are quite similar to those of the damped oscillatory eigenmode (not
shown) of the linearized system, whose period and decay time are $51$
and $44$ years, respectively.  Therefore, although the initial
perturbation projects onto various eigenmodes, the system's response
is dominated by the damped oscillatory eigenmode of the linearized
system after a relatively short adjustment time.

The relative phases of the oscillation between the different boxes can
be understood as time delays for the advection of the anomalous water
through the boxes.  The amount of delay for the advection is directly
related to the size of the box.  The relative amplitude of the
oscillation within a box is also related to the sizes of the boxes,
with the largest box (box 3) exhibiting the smallest amplitude.
Additionally, the surface temperature anomalies, as they are damped by
the atmospheric restoring, are reduced in amplitude relative to the
surface salinity anomalies which are not damped under the fixed
$(E-P)$ fluxes.  The temperature anomalies are hence slightly
subdominant to the salinity anomalies in their contribution to the
anomalous circulation.  Nevertheless, temperature effects play an
important and non-negligible part in the physical mechanism
responsible for the model's oscillations, as explained in the
following.

The temperature and salinity perturbations within each box determine
the circulation anomaly passing through all boxes.  The circulation
anomaly in turn creates a feedback to the temperature and salinity in
the individual boxes.  Note that the results reported throughout this
paper are generated by the full nonlinear conservation equations
(\ref{eq:T4prime})--(\ref{eq:S4prime}).  However, since the model is
exhibiting a very linear response, it is useful for understanding the
physical mechanism of this damped oscillation to follow a cycle with
reference made to the linear equations
(\ref{eq:T4prime})--(\ref{eq:S4prime}).

We begin with the growing phase of a positive circulation anomaly
(e.g., year 125--130 in Fig.\ \ref{fig:combined_plots}), where the
positive circulation anomaly $U'$ advects the mean high salinity
$\overline{S}_{1}$ and mean high temperature $\overline{T}_{1}$ water
from the southern surface box (box 1) to the northern surface box (box
2) which drives up the salt and temperature anomalies in the north
[see in particular the first terms of the right hand side of eqns.\
(\ref{eq:T2prime}) and (\ref{eq:S2prime})].  A comparison of Figs.\
\ref{fig:combined_plots}A and \ref{fig:combined_plots}B indicates that
the growth in dimensionless salinity anomaly $\beta S_{2}'$ dominates
the dimensionless temperature anomaly $\alpha T_{2}'$ during this
portion of the oscillation because of the effects of the atmospheric
restoring term ($-\gamma_{T} T_{2}'$) on the warm water advected from
box 1.  This process drives up the northern density which amplifies
the growing circulation anomaly and thus represents a positive
feedback.  Note that during this buildup of salty water in the north,
a fresh cold anomaly develops in the southern upper box 1.  So far the
discussion parallels the thermohaline instability mechanism under
mixed boundary conditions described by Walin (1985) and Marotzke et
al.\ (1988), in which the temperature variations are neglected.

The positive feedback of added salinity in the north eventually
yields, within a few years, to a negative feedback due to the warm
water also advected northward which reduces the growth of the northern
density and therefore the circulation anomaly as well. The negative
feedback from temperature eventually causes the circulation anomaly
and the northern density to reach a maximum.  The peaking of the
northern density causes the circulation anomaly to likewise reach a
maximum and thereafter begin the decreasing portion of its
oscillation.  The extremum reached by the density in a box corresponds
to the interchange in dominance between the competing terms in the
linearized equations (\ref{eq:T1prime})--(\ref{eq:S4prime}).  In
particular, when the circulation anomaly has weakened sufficiently due
to the temperature feedback, the advection of the anomalous
temperature and salinity gradients by the mean circulation [i.e.,
$\overline{U}(S_{1}'-S_{2}')$ and $\overline{U}(T_{1}'-T_{2}')$]
dominates the advection of the mean gradients by the circulation
anomaly [ i.e., $U'(\overline{S}_{1}-\overline{S}_{2})$ and
$U'(\overline{T}_{1}-\overline{T}_{2})$]; cf.\ equs.\
(\ref{eq:T2prime}) and (\ref{eq:S2prime}).  This interchange in
dominance has a stabilizing effect and causes both the temperature and
salinity of box 2 to continue decreasing.  The atmospheric restoring
term $-\gamma_{T}T_{2}'$ also cools the warm water pool formed in box
2, again reducing the temperature perturbation towards zero.

Once the circulation anomaly approaches the zero crossing point, the
perturbation temperature and salinity in the northern boxes are also
approaching the zero crossing point.  Because of the phase lag between
the north and south, there is still a significant fresh cold anomaly
in the southern box 1 which is now advected by the mean flow to the
north.  This advection forces the zero crossing of the temperature and
salinity perturbations in the north and creates a negative density
anomaly.  This density anomaly induces a negative circulation anomaly;
i.e., it weakens the total circulation.  The same cycle described
above now repeats but with the opposite temperature, salinity, density
and circulation anomalies.  In this opposite portion of the
oscillation, the negative circulation anomaly, which is initially
strengthened in absolute value by the positive feedback created by the
negative salinity anomaly in the north, will reach an extremum when
the negative feedback due to the negative temperature anomaly in the
north kicks in.  The circulation anomaly thereafter increases towards
zero again, thus completing one full cycle.

It should be re-emphasized that although temperature anomalies are
slightly subdominant to salinity anomalies in their contribution to
the box model's circulation anomaly, their effects are non-negligible
in the physical mechanism presented here.  For example, in the absence
of the negative feedback from temperature, a thermally dominant steady
state is unstable to perturbations.  This behaviour, which is clear
from the mechanism previously described, can be verified
mathematically by considering the response of the model when the
temperature feedback is suppressed.  In particular, a linear stability
analysis for a fixed temperature perturbation (all temperatures are
held to their steady state values) yields exponentially growing
non-oscillatory eigenmodes signaling an instability.  Conversely,
perturbations to the model in which the salinity feedback is
suppressed are exponentially damped and non-oscillatory due to the
absence of a positive feedback in the system.

Therefore, the linear mechanism for the oscillations relies on three
main factors: a positive feedback from salinity that creates a growth
in the absolute value of the salinity, temperature and circulation
anomalies; a negative feedback from temperature, following a few years
after the positive feedback from salinity, that dampens the growth
caused by the salinity feedback allowing the anomalies to reach a
local extremum; and a time delay between the temperature and salinity
in the southern and northern boxes that allows for the zero crossing
of the oscillating perturbations.  Finally, the overall damping of the
oscillations seen in Fig.\ \ref{fig:combined_plots}\ reflects the
dominance of the temperature feedbacks for the perturbations about
this particular steady state.

An important element of the oscillation mechanism described above, as
well as that discussed by D93 for the coupled model, is the phase lag
between the temperature and salinity within a particular region of the
respective models.  This lag in the box model is a direct result of
the different boundary conditions for the temperature and salinity.
When both temperature and salinity have the same boundary conditions,
whether a fixed $(E-P)$ and fixed heat flux or restoring conditions
with the same restoring times, the temperature and salinity equations
can be reduced to a single density equation when a linear equation of
state is used.  Model variability will therefore have either a zero or
180 degree phase difference between the anomalous temperature and
salinity fields thus making the variability density driven rather than
thermohaline (i.e., temperature {\em and} salinity) driven.  Breaking
the symmetry between temperature and salinity forcing, as through
mixed boundary conditions, allows the two fields to no longer have
such a constrained phase relationship thus enabling the existence of
the linear thermohaline oscillations described here.

It is useful to compare the oscillations described here to the
Howard-Malkus loop oscillations described by Welander (1986) and more
recently by Winton and Sarachik (1993).  Since both oscillators are
advective in nature, their period is related to the time for the
advection of anomalous fields through the respective systems. In the
loop systems, self-sustained nonlinear oscillations are possible due
to the existence of unstable equilibria which result in a limit cycle
mechanism.  The oscillations described here result from linear
perturbations about a stable equilibrium and are hence exponentially
decaying and not self-sustaining.

\section{Stochastically forced thermohaline oscillations}
\label{section:Random_forcing}

In this section, we consider the response of the box model to
stochastic forcing on the surface boxes under mixed boundary
conditions.  This forcing is meant to approximate the effects of
synoptic scale atmospheric variability on the air-sea fluxes present
in the coupled model of D93.  The numerical realization of the
stochastic forcing follows that described by Kloeden and Platen (1992)
(for a mathematical discussion) and Penland (1989) (for meteorological
examples).  The response of the fully nonlinear box model to a single
perturbation in Section \ref{section:Linear_oscillations} was
understood as being predominantly the response of the model's damped
oscillatory eigenmode from the linearized system.  This linear
interpretation will also hold in the following for the model when
driven with a moderate amount of stochastic forcing.

\subsection{The stochastic forcing}
\label{subsection:stochastics}

To motivate the form of the stochastic forcing acting on the surface
boxes, it is useful to consider the results of D93 for the regression
of the zonally averaged anomalous heat and salinity budgets in the
northern sinking region of their model against the anomalous THC index
(Figs.\ 14 and 15 of D93, respectively).  The regression coefficient
at all lag times for the anomalous surface ($E-P$) flux is very small
relative to that for the salinity transported by the meridional
oceanic circulation from the south.  While this does not imply a small
($E-P$) flux over the synoptic time scale of the coupled model's
atmospheric variability, it does indicate that the variability in the
surface salinity forcing over the model's sinking region has a
negligible contribution to the interdecadal THC variability.  We
therefore choose to set the anomalous ($E-P$) forcing to zero in the
following box model study.

The regression coefficients for the coupled model's anomalous surface
heat flux are of similar magnitude to those from the meridional heat
transported by the oceanic circulation from the south.  Therefore, we
expect the variability in the surface heating to be important for the
THC variability and we therefore use an anomalous stochastic heat
forcing in the following.  Furthermore, since the interdecadal time
scale of interest here is so much longer than the typical synoptic
atmospheric weather phenomena, the stochastic heating will be modeled
as an additive white noise heating.  Experiments with nonzero
auto-correlation (not shown) give qualitatively similar results to
those reported here for the white noise heating.

 To include the stochastic heating, the surface box temperature
equations (\ref{eq:T1}) and (\ref{eq:T2}) are modified to the
following form with additive white noise forcing
 \begin{eqnarray}
  \dot{T_{1}} &=& {U \over \delta V}(T_{3} - T_{1}) +
   \gamma_{T}(T_{1}^{*} - T_{1}) +
k\gamma_{T}(T_{1}^{*}-\overline{T}_{1})\psi_{1}
\label{eq:T1_noise}\\
  \dot{T_{2}} &=& {U \over \epsilon \delta V}(T_{1} - T_{2}) +
  \gamma_{T}(T_{2}^{*} - T_{2}) +
k\gamma_{T}(T_{2}^{*}-\overline{T}_{2})\psi_{2}.
 \end{eqnarray}
 In these expressions, $k$ is an adjustable noise level scaling and
$\psi_{1,2}$ are independent Gaussian distributed, unit variance white
noise processes.  A value of $k=.08$ is used in the following
experiments, which indicates a white noise forcing with standard
deviation $8 \%$ the magnitude of the steady state heat flux.  The
white noise forcing is added every time step (365 time steps per year)
during the integration.  The remaining box model parameters are those
given in Section \ref{section:Linear_oscillations}.

Both the box model and the coupled GCM contain rapidly de-correlating
processes that are not of direct interest in the present study.
Therefore, to focus on the interdecadal response of the models, a $10$
year low pass filter is applied to the yearly averaged signals
generated from the box model.  The coupled model contains a trend
which is roughly linear over the $200$ years of integration considered
here.  As described in D93, this trend is removed and then a $10$-year
low pass filter is applied to the yearly averaged fields.  The
subsequent analyses presented here consider only the low pass
anomalous fields generated from both models.

\subsection{Box model--coupled GCM comparison}
\label{subsection:box_gcm}

In this and the subsequent section, we compare part of the analysis of
a $200$ year simulation from the coupled model of D93 to a $200$ year
simulation from the box model.  The statistics presented below for the
box model were validated by using significantly longer time series.
Since the ocean portion of the coupled model did not make a transition
to a new equilibrium during the millennium integrated, the ocean will
be considered to be in a stable climate regime.  Likewise, the box
model will be put in a thermally dominant regime such as that
considered in Section \ref{section:Linear_oscillations}.  The
auto-correlation function of both the THC index, defined as the
maximum yearly averaged transport within the coupled model's North
Atlantic region, and the box model's circulation anomaly are used as a
diagnostic for determining the precise regime of the box model roughly
corresponding to the coupled model's behaviour.

\begin{figure}[htbp]
\centerline{\psfig{figure=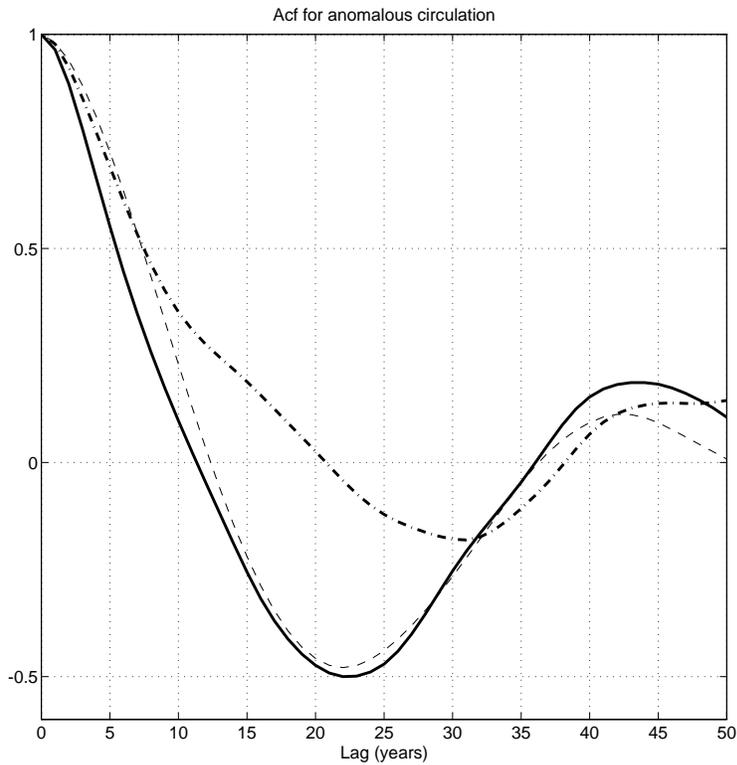,height=4.0in}}
\caption{
   \baselineskip 3ex
   Auto-correlation functions of the 
   anomalous THC for the box model 
   (dashed and dot-dashed lines) and the anomalous THC index for the 
   coupled model (solid line).
   The minima at lags $\approx \pm 25$ years for the coupled model and 
   the dashed line box model indicate the time scale for a
   typical oscillation.  This time scale for the box model is 
   determined by the period ($\approx 50$ years) 
   of the damped oscillatory eigenmode.  The dot-dashed line is for the box
   model in a highly damped oscillatory regime (i.e., ratio of the decay time to
   period  is small). }
\label{fig:auto_corr}
\end{figure}

Fig.\ \ref{fig:auto_corr} shows the auto-correlation function for the
THC index from the coupled model (solid line) as well as for the
circulation anomaly from the stochastically forced box model.  For the
box model, we present the circulation's auto-correlation function
resulting from two different salinity forcings.  The dot-dashed line
is the auto-correlation for the model placed in a highly damped
oscillatory regime.  For this experiment, the salinity restoring time,
used to calculate the steady state before switching to mixed boundary
conditions and the subsequent stochastic forcing, is set to
$\gamma_{S}^{-1} = 900$ days.  The resulting salinity forcing for the
mixed boundary condition integration is relatively weak which results
in a highly damped oscillatory response.  The resulting steady state
circulation $\overline{U} = 19.6$ Sv.  The damped oscillatory
eigenmode of the system linearized about this steady state has period
and e-folding time of $62$ and $9.8$ years, respectively.  The dashed
line in Fig.\ \ref{fig:auto_corr}, which closely follows the
auto-correlation function from the coupled model, is the result of
using the salinity restoring coefficient of $\gamma_{S}^{-1} = 300$
days which was considered in Sections \ref{subsection:parameters} and
\ref{section:Linear_oscillations}.  Motivated by the close
correspondence between the auto-correlation functions of the coupled
model and that of the box model spun up with $\gamma_{S}^{-1} = 300$
days, we further study the variability of the mixed boundary condition
box model spun up with this salinity forcing and compare to the
coupled model results.

\begin{figure}[htbp]
\centerline{\psfig{figure=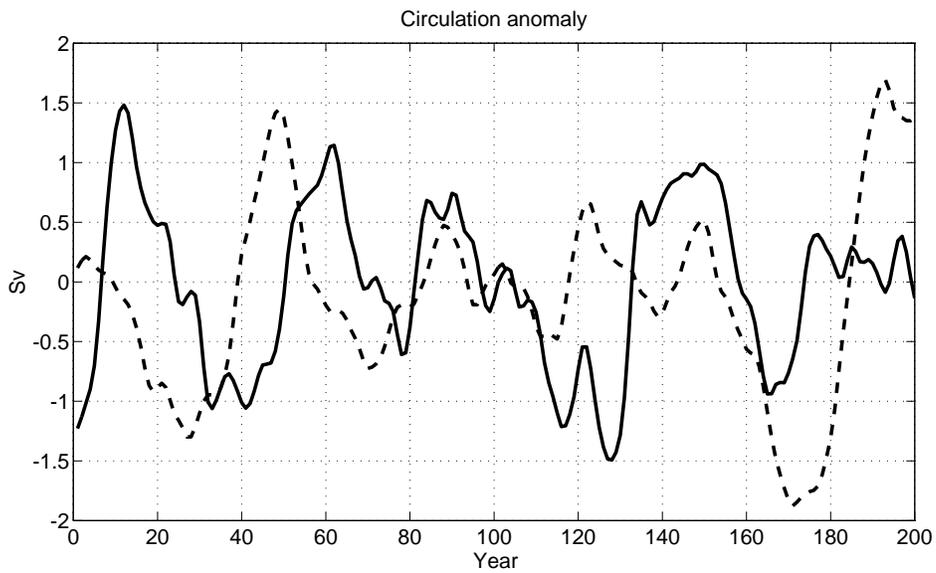,height=3.0in}}
\caption{
   \baselineskip 3ex
   Yearly averages of the 
   circulation anomaly in the box model
   with stochastic heat forcing on the surface boxes (dashed line) 
   and the linearly detrended THC index for the first $200$ years of the 
   coupled model of D93 (solid line).   
   The circulation of the underlying thermally dominant 
   mean state is $16.9$ Sv for the box model and its standard 
   deviation is $0.77$ Sv.  The mean state for the coupled 
   model is $18.3$ Sv and the standard deviation is $0.68$ Sv.
   These time series have auto-correlation functions shown in 
   Fig.\ \protect\ref{fig:auto_corr}. }    
\label{fig:circulation}
\end{figure}

The time series for the annually averaged anomalous meridional
circulation in the box model driven by stochastic forcing is shown in
Fig.~\ref{fig:circulation}.  The power spectra for this time series
(not shown) has power broadly centered at a period of $50$ years,
reflecting the contribution of the damped oscillatory eigenmode for
the variability in the stochastically forced system.  The
corresponding time series for the thermohaline index from D93 is also
shown in Fig.\ \ref{fig:circulation}.  The power spectrum for this
time series, shown in Fig.\ 5 of D93, has power broadly centered
around $40$--$50$ years.

\begin{figure}[htbp]
\centerline{\psfig{figure=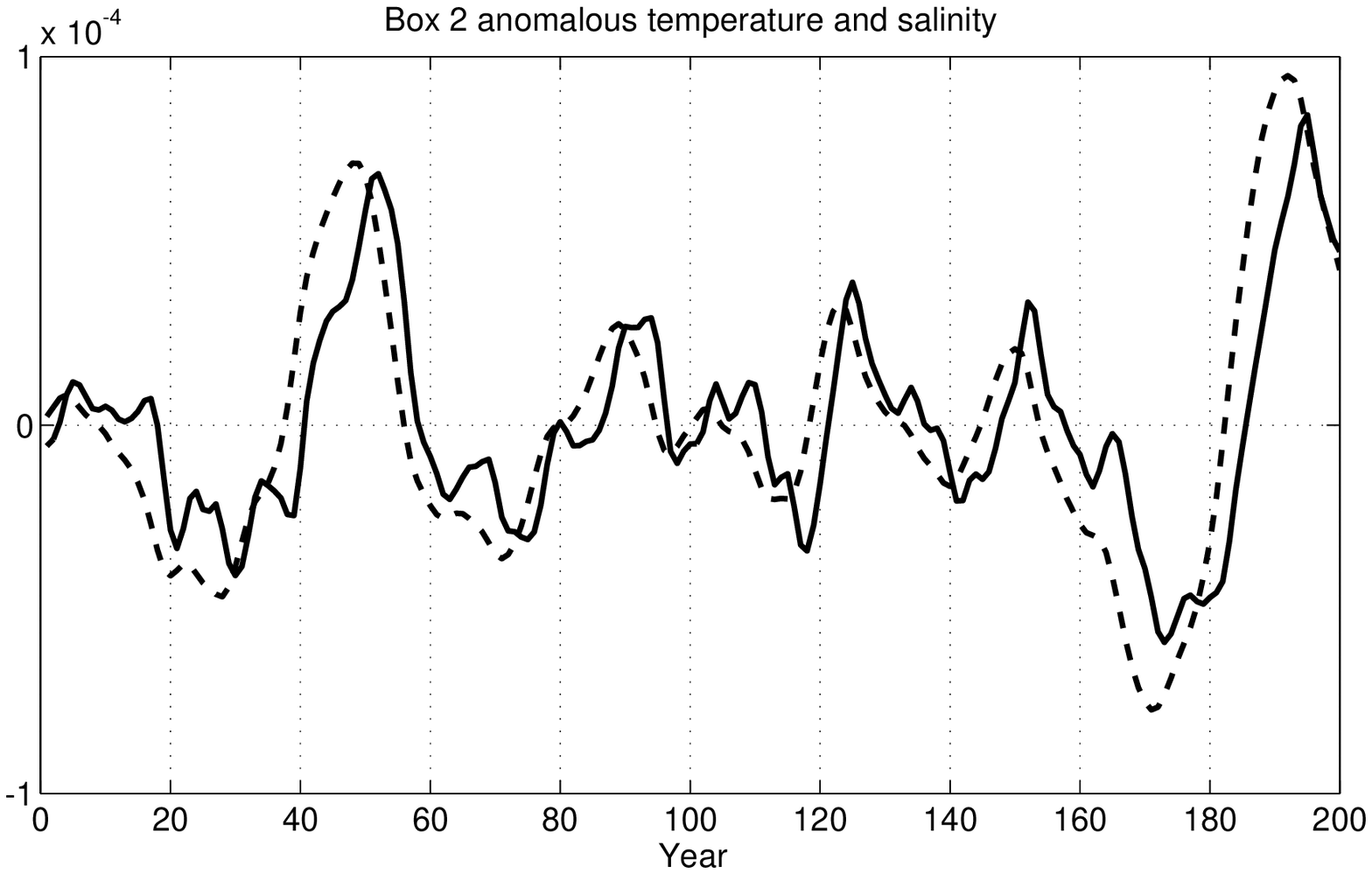,height=3.0in}}
\centerline{\psfig{figure=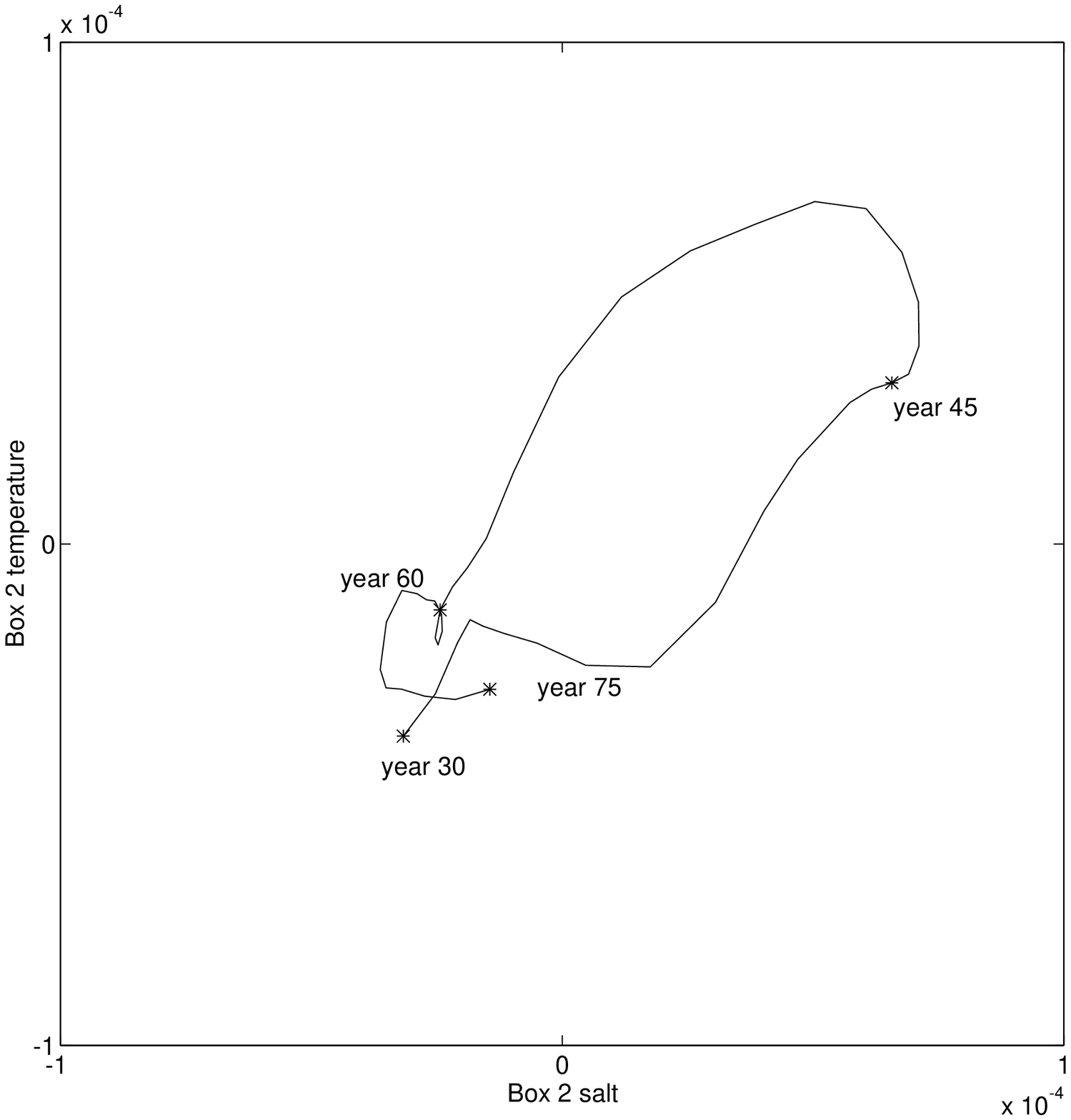,height=4.0in}}
\caption{
   \baselineskip 3ex
   (A) Dimensionless temperature anomaly
   $\alpha T_{2}'$ (solid line) and salinity anomaly
   $\beta S_{2}'$ (dashed line)  
   for the northern surface box (box 2) in the  
   box model with surface stochastic forcing.  Note the phase relation
   (salt leads temperature) indicative of the oscillating mode shown
   in Fig.\ $2$.
   (B) ($\beta S_{2}',\alpha T_{2}')$  plane for years 30-70 
   of (A) with selected years indicated.  
   The trajectory is in the counterclockwise direction since
   salt leads temperature. }
\label{fig:northern_temp_salt}
\end{figure}

The time series for the temperature and salinity in the northern top
box (box $2$) in the box model are shown in
Fig.~\ref{fig:northern_temp_salt}A.  The response of the remaining
larger boxes (not shown) are smoother and of smaller amplitude than
the surface boxes reflecting an integrating property of the box model
in its response to surface stochastic forcing.  The phase relation
(salt leads temperature), also present in the oscillating eigenmode
shown in Fig.\ \ref{fig:combined_plots}, reflects the dominance of
this mode in determining the response of the stochastically forced
system.  The phase relations are further illustrated in the ($\beta
S_{2}',\alpha T_{2}')$ plane shown in
Fig.\ref{fig:northern_temp_salt}B.  As salt leads temperature, the
roughly circular trajectory traced out in this diagram is in the
counterclockwise direction.

\subsection{Linear regression analysis}
\label{subsection:regressions}

In order to explore the physical mechanism of the interdecadal
variability using time series from the coupled model, D93 computed the
linear regression coefficients between the time series of selected
model fields in the ``sinking region'' ($52^{\circ}$N to
$72^{\circ}$N) of the North Atlantic portion of their model against
the time series for the meridional circulation anomaly in the same
region.  This analysis provides information concerning the phase
relations between the circulation anomaly and the regressed field;
information which may be obscured in the individual noisy time series.
For comparing these results with those of the box model, we employ the
same regression analysis here.

The linear regressions involve the least squares fit of a linear
relation between a selected anomalous field $F(t)$ (e.g., salinity,
heat flux, etc.) and the circulation anomaly $U'(t)$ over lagged
portions of their respective time series of length $T (200$ years);
i.e., we compute a regression coefficient at each lag $\tau \ge 0$ by
minimizing the quantity
 \begin{equation}
 \sum_{\tau \le t \le T}[ F(t) - a(\tau)U'(t-\tau) - b(\tau) ]^{2}.
 \label{eq:regress}
 \end{equation}
 For lags $\tau \le 0$, the sum runs from $|\tau|$ to $T$ and the
arguments $F(t+\tau)$ and $U(t)$ are used.  The regression coefficient
$a(\tau)$ will be presented for lags $-30 \le \tau \le 30$ years.
Note that negative lags represent times prior to the maxima of the THC
and positive lags are times subsequent to the maxima.

\begin{figure}[htbp]
\centerline{\psfig{figure=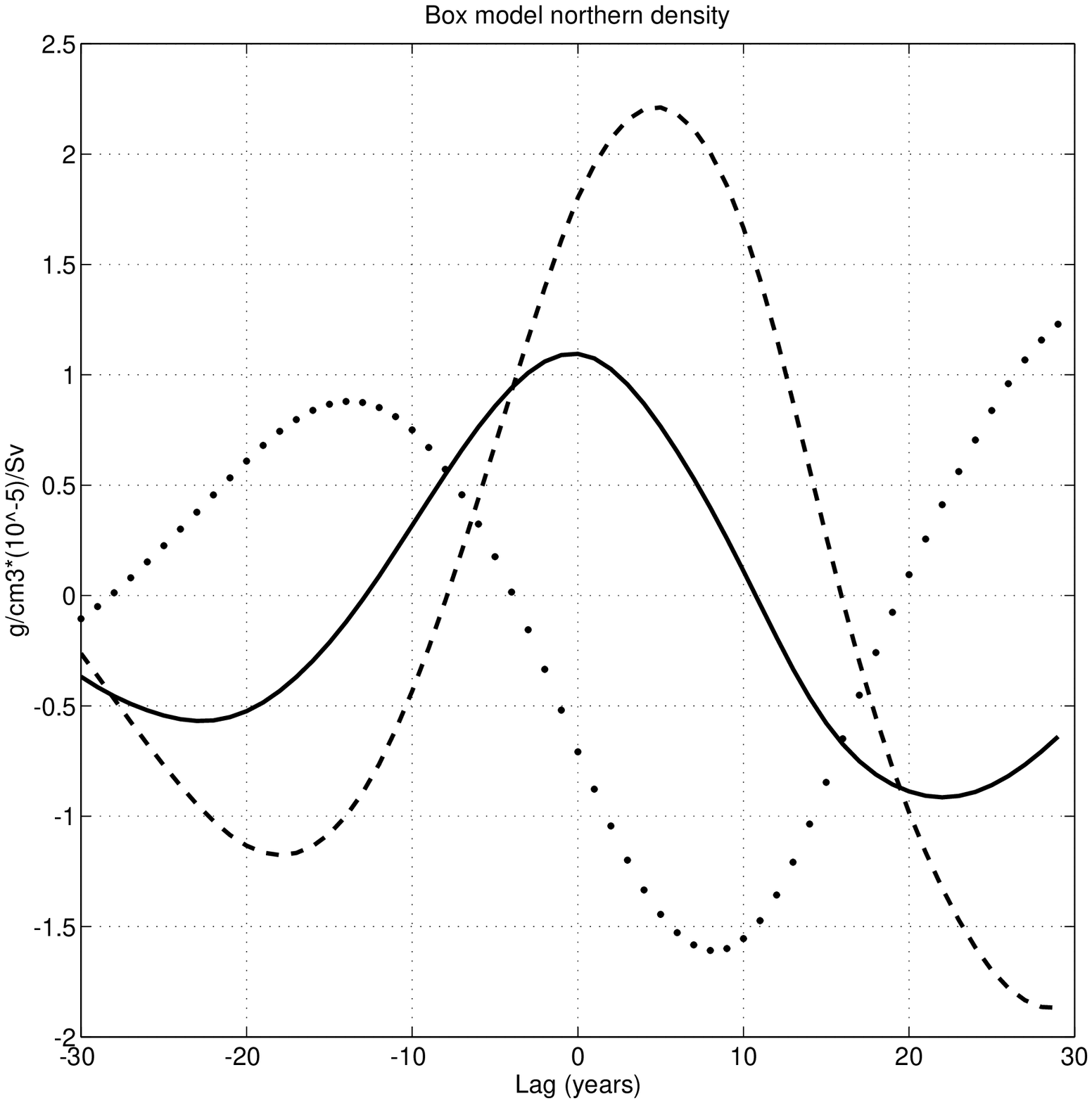,height=3.75in}}
\centerline{\psfig{figure=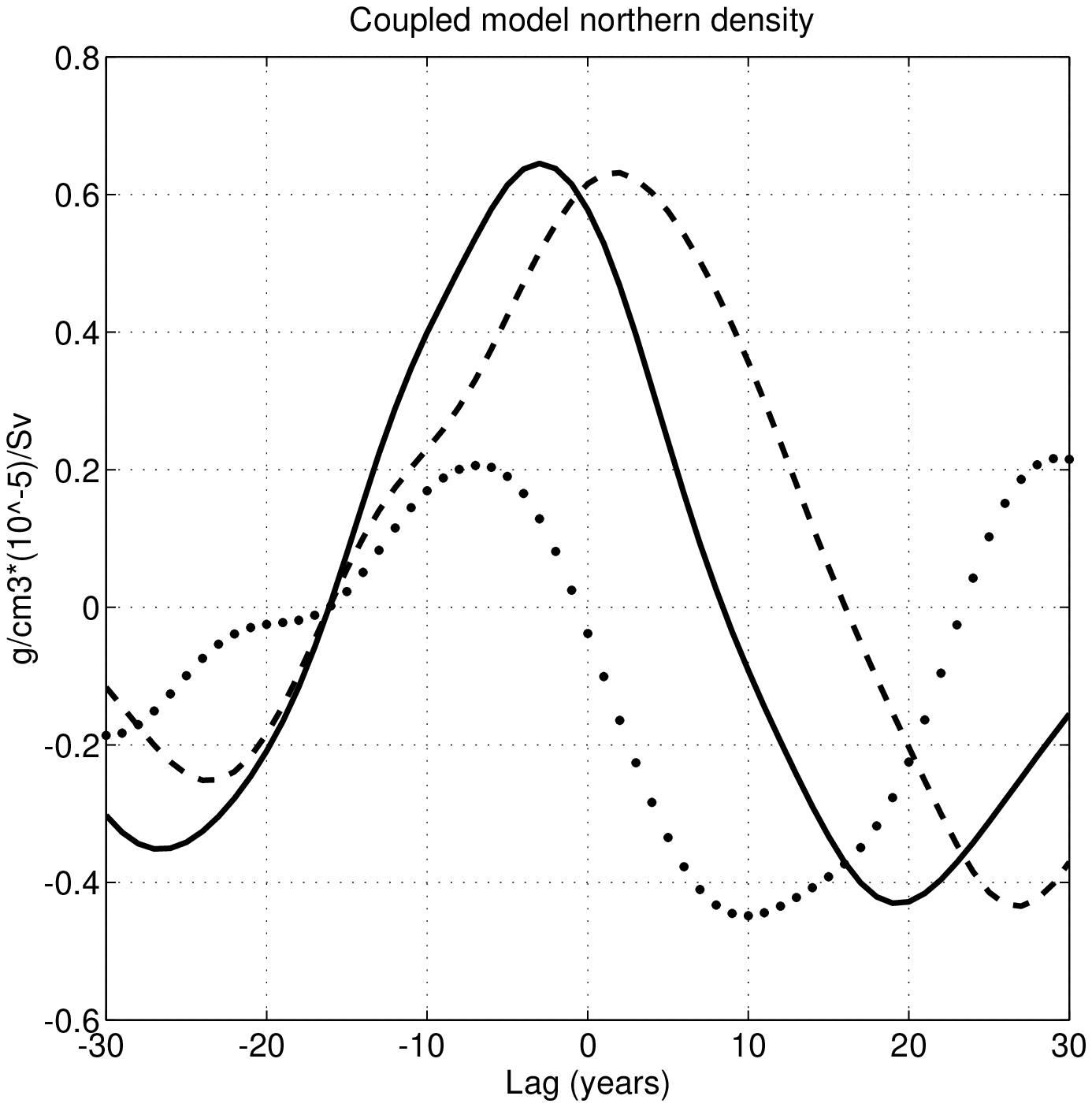,height=3.75in}}
\caption{
   \baselineskip 3ex
   (A) Linear regression of the 
   anomalous density in the northern boxes
   against the filtered anomalous thermohaline circulation. 
   The double-dashed line is for
   $\rho_{salt} = \rho_{0} \beta(S_{4}' + \delta S_{2}')/(1+\delta)$, 
   the dotted line is for 
   $\rho_{temp} = -\rho_{0} \alpha(T_{4}' + \delta T_{2}')/(1+\delta)$,
   and the solid line is for the total density 
   $\rho_{total} = \rho_{salt}+\rho_{temp}$.
   (B) Corresponding regressions for the coupled model (Fig.\ $8$ of D93).}
\label{fig:density_regress}
\end{figure}

Fig.~\ref{fig:density_regress}A shows the regression coefficient of
the box model density anomaly spatially averaged over the northern
boxes, $(\delta \rho_{2} + \rho_{4})/(1+\delta)$, and its temperature
and salinity contributions.  The density peaks at lag zero indicating
that the northern density is in phase with the circulation, a fact
consistent with the damped oscillatory eigenmode dominating the
oscillations as can be deduced from the time series in Fig.\
\ref{fig:combined_plots}.  Fig.~\ref{fig:density_regress}B shows the
corresponding regression from the coupled model for its sinking
region.

Comparison of the two regressions reveals some of the similarities and
differences between the two models' THC variability.  Notably, the box
model salt and temperature contributions to density have similar
relative phases as in the coupled model.  Additionally, both models
indicate that salinity is the dominant contributor to the density
anomaly, and hence to the overall circulation anomaly.  As mentioned
previously, it is the damping of surface temperature anomalies which
reduces their response relative to salinity anomalies in the box
model.  The magnitudes of the regression coefficients are larger in
the box model than in the coupled model.  This quantitative difference
is, however, not unexpected considering the idealized nature of the
box model.  The phase difference between salt and temperature
contributions to the density as seen in the box model is larger than
that of the coupled model.

\begin{figure}[htbp]
\centerline{\psfig{figure=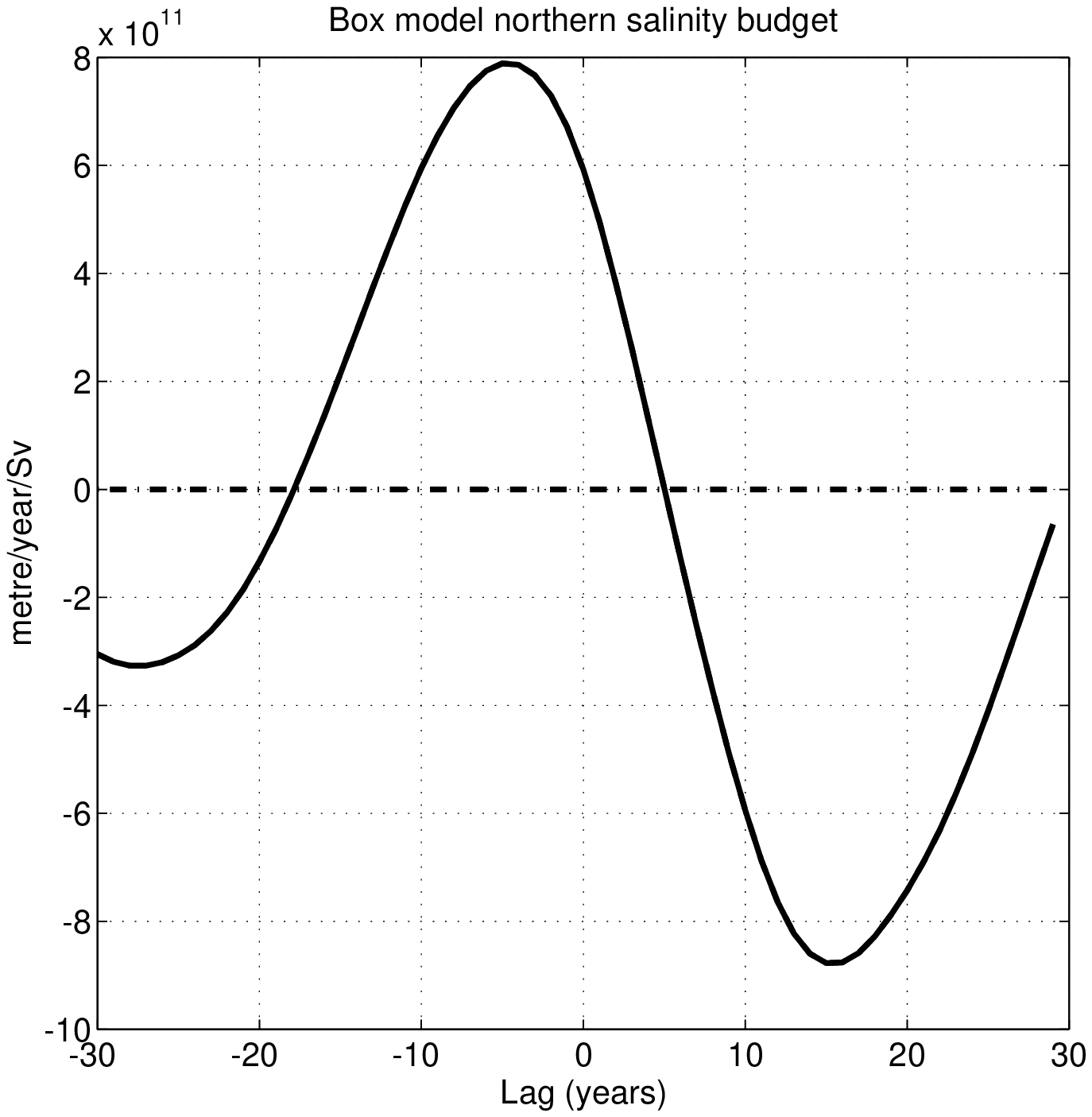,height=3.75in}}
\centerline{\psfig{figure=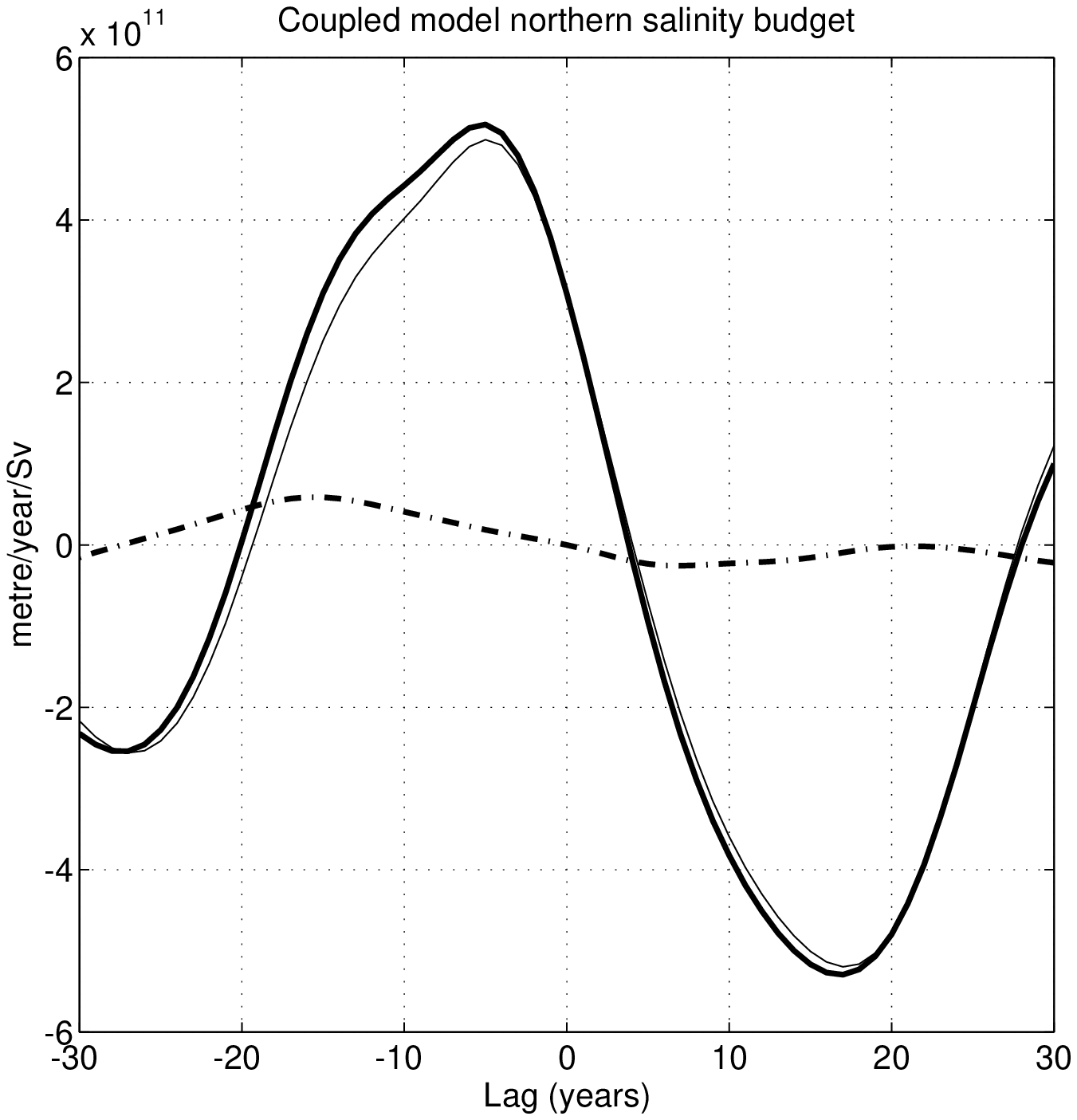,height=3.75in}}
\caption{
   \baselineskip 3ex
   (A) Linear regression of the anomalous salt budget over the northern 
   boxes against the anomalous thermohaline circulation. The
   dashed line is for
   the surface salt transport, which has a zero anomaly in this experiment, 
   and the solid line is for meridional oceanic salt transported 
   into the north.  This anomalous salt transport is given by 
   $U(S_{1}-S_{4})/S_{0} -
   \overline{U}(\overline{S}_{1}-\overline{S}_{4})/S_{0}$, 
   where the reference salinity $S_{0}=35$ psu.
   (B) The corresponding regression coefficients 
   for the coupled model (Fig.\ $15$ of D93, without the transport from
   the north) with the dashed
   line for surface transport, the thin solid line for the oceanic
   meridional transport, and the thick solid line their sum.  }
\label{fig:salt_regress}
\end{figure}

\begin{figure}[htbp]
\centerline{\psfig{figure=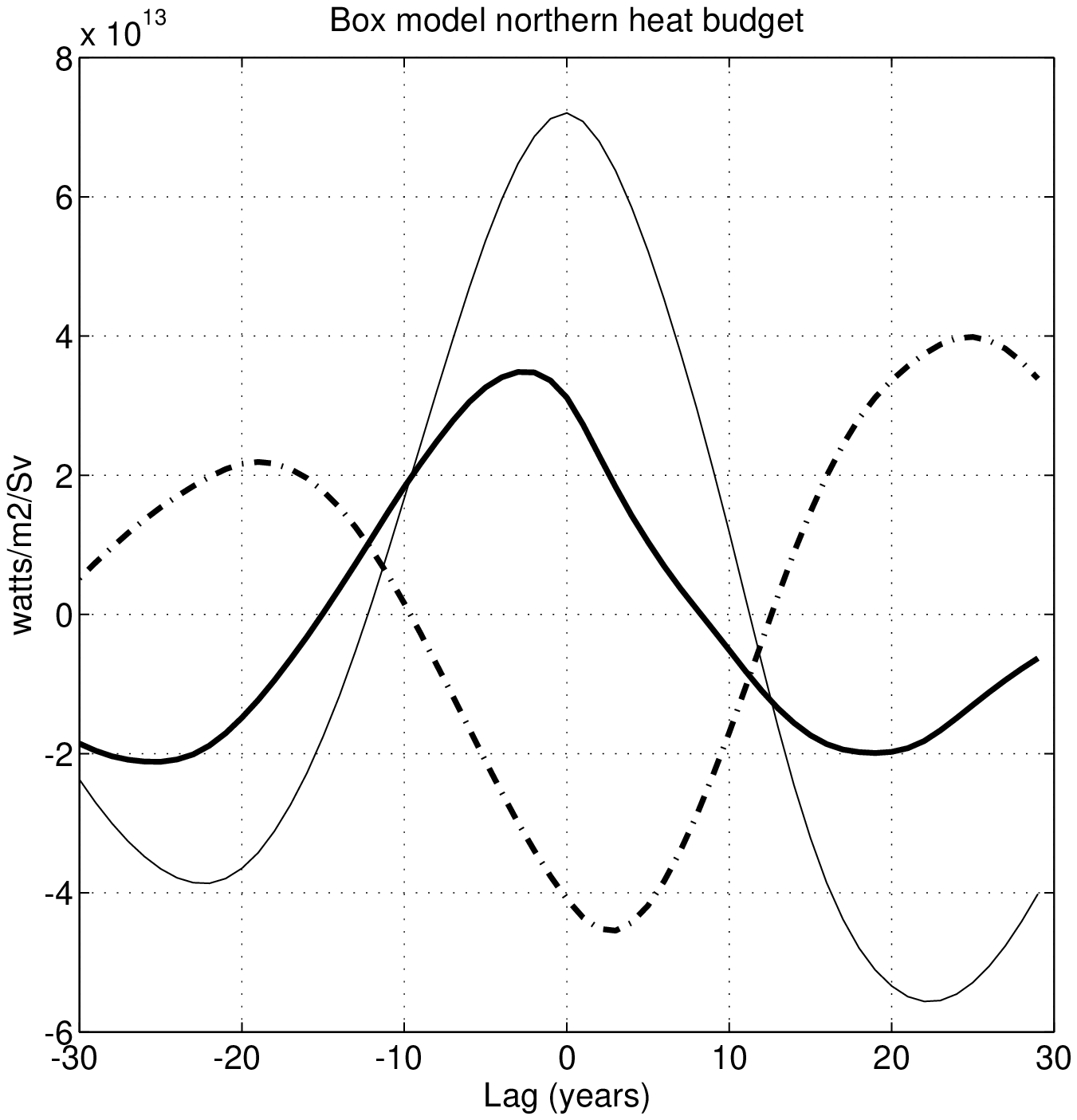,height=3.75in}}
\centerline{\psfig{figure=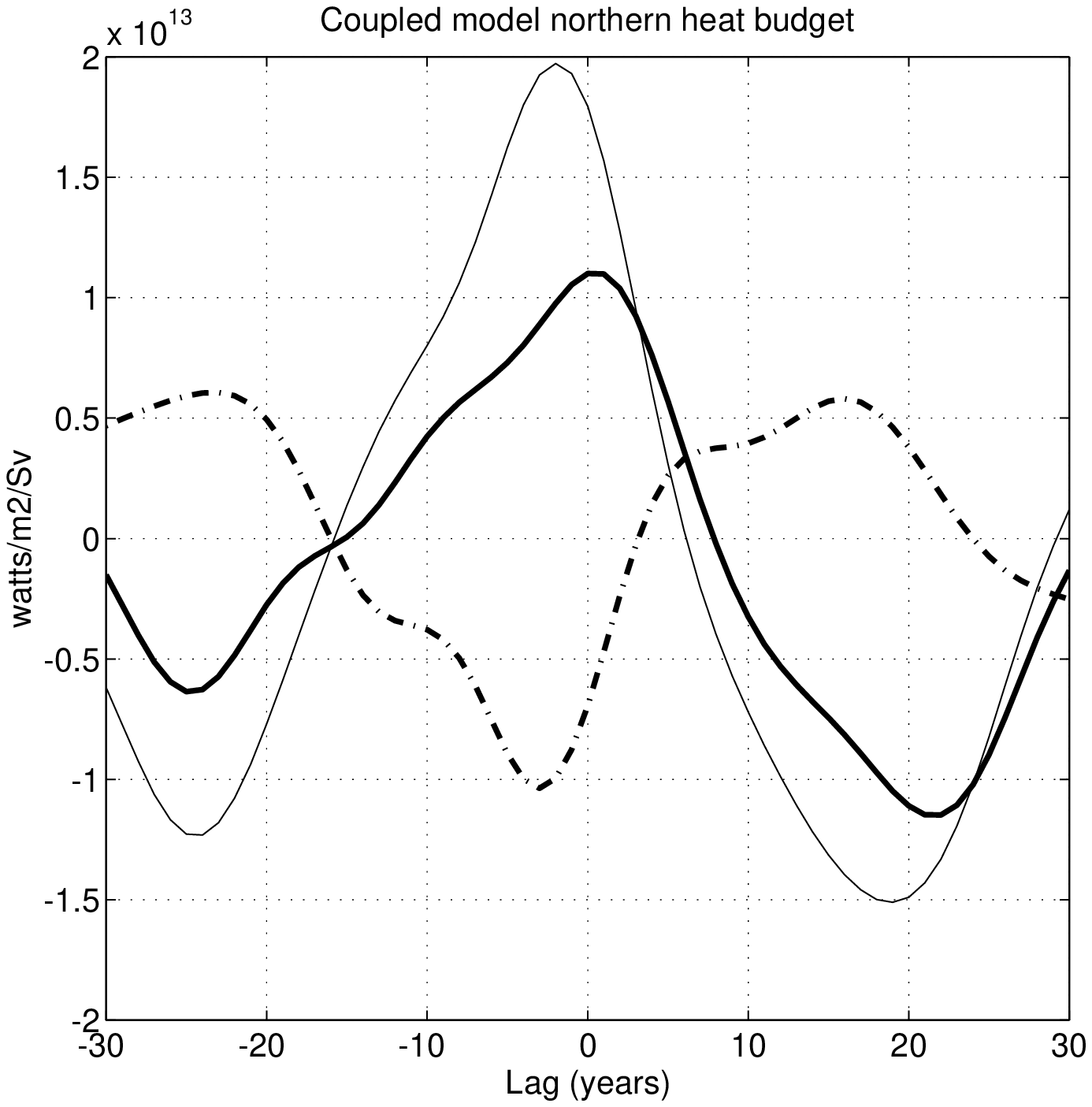,height=3.75in}}
\caption{
   \baselineskip 3ex
   (A) Linear regression of the anomalous heat budget over the northern 
   boxes against the anomalous THC . The
   dashed line is for the surface transport,  
   the thin solid line is for the oceanic meridional transport,
   and the thick solid line is for their sum.
   The anomalous surface heat contribution is given by 
   $ (\epsilon \delta V) 
   \rho_{0} C_{p} \gamma_{T}[(T_{2}^{*}-T_{2})-(T_{2}^{*}-\overline{T}_{2})]$
   and the anomalous meridional heat transport by 
   $\rho_{0} C_{p}[ U(T_{1}-T_{4}) -
   \overline{U}(\overline{T}_{1}-\overline{T}_{4})]$, 
   with $\rho_{0} = 1027 {\mbox kg}/{\mbox m}^{3}$ and 
   $C_{p} = 4000 {\mbox J}/{\mbox kg}^{\circ}{\mbox C}$.
   (B) Corresponding regression coefficients 
   for the coupled model (Fig.\ $14$ of D93), with the thin solid 
   line for oceanic meridional transport,  the dashed line for the surface
   transport, and the thick solid line for their sum.  }
\label{fig:heat_regress}
\end{figure}

Fig.~\ref{fig:salt_regress}A shows the regression coefficients for the
anomalous salinity budget transported into the northern box through
the surface as well as that transported by the ocean from the south.
Fig.~\ref{fig:salt_regress}B shows the corresponding regression for
the coupled model.  Note that in D93, a southward oceanic meridional
transport into the sinking region from the extreme northern sector of
the model was also included in their regression analysis.  This
transport contributes a negligible amount to the regression
coefficients and is therefore not reproduced here.  The box model
$(E-P)$ budget of Fig.\ \ref{fig:salt_regress}A is implied from the
salinity transports divided by a reference salinity of $S_{0}=35$ psu.
Recall that the anomalous surface salinity flux in the box model is
set to zero.  Figs.~\ref{fig:heat_regress}A,B present the
corresponding regression coefficients for the heat budget in the two
models.  The sign conventions are such that a positive regression
coefficient implies a positive anomalous transport into the northern
portion of the model ocean (i.e., positive salinity and positive heat)
for a positive anomalous circulation.

In the coupled model, an anomalous gyre transport plays an important
role in advecting a positive salinity anomaly into the north previous
to the THC maximum.  Comparison of Fig.~\ref{fig:salt_regress}A with
the regression for the coupled model in Fig.~\ref{fig:salt_regress}B
indicates that without the gyre, the box model also brings a positive
salinity anomaly into the north prior to the maximum of the THC, hence
emulating the transport exhibited in the coupled model.  The
similarity of the variability in the two models might suggest that the
gyre transport may be essential for establishing the detailed
structure of the variability, yet perhaps the existence of the
oscillation could be explained without the gyre effect.  We note that
Greatbatch and Zhang (1993) found a gyre effect similar to D93 induced
by meridional oscillations based on a thermal only mechanism (i.e.,
the salinity was fixed in this model so that its oscillation mechanism
is not thermohaline but rather single component).

Both the box and coupled models exhibit the same respective order of
magnitude for the contributions of the surface and meridional heat
transports into the northern region.  This result is an important
element of the box model oscillation and is associated with the
restoring conditions placed on the surface temperature.  Recall that a
different perspective has been taken in the study of Greatbatch and
Zhang (1994) and Cai et al.\ (1994) who study the THC oscillations in
a model forced with a fixed surface heat flux; i.e., a zero anomalous
surface heating.  Their justification for considering this forcing was
based on the dominance of the anomalous meridional heat transport over
the anomalous surface heat transport in the coupled model THC
variability, as revealed by the regression plot in
Fig.~\ref{fig:heat_regress}B.  We chose the conventional mixed
boundary conditions formulation for the box model study to emulate the
non-trivial phase relations between salinity and temperature and the
THC seen in the coupled model.  Such relations are not found in models
forced with fixed heat and salinity fluxes.  Additionally, both the
box and coupled models have their surface and meridional heat
transport components roughly $180^{\circ}$ out of phase.  This
similarity indicates that both models respond to the increasing
circulation by bringing more warm water northward, with the peak
happening near the peak of the THC.  Note that the qualitative
agreement between the models' salinity regressions shown in
Figs.~\ref{fig:salt_regress}A,B motivated the neglect of an anomalous
surface salinity flux in the box model study.

\section{Discussion}
\label{section:discussion}

The mechanism operating in the box model oscillations is advective.
Consequently, the period of the oscillations was found to be sensitive
to the volume of the ``sinking region'' of the box model (boxes 2 and
4).  If a similar dependence exists in the coupled model, this result
may suggest that a more realistic higher resolution ocean model than
that of D93, in which the extent of the sinking region may change,
could result in a different time scale for the variability than that
found in D93.

It is useful to contrast the roles played by temperature and salinity
during the oscillations with their roles in establishing the
underlying mean circulation.  For the mean circulation, cold
temperatures in the north and warm temperatures in the south act to
set up a circulation with sinking in the north.  Conversely, the
buoyancy torque contributed by salinity acts to brake the mean
thermally dominant circulation.  If strong enough, the salinity torque
causes the system to make a transition to the saline dominant
circulation.  For the thermohaline oscillations about a stable
thermally dominant circulation, temperature effects act to move the
anomalies back towards zero and thus acts as a negative feedback to
the oscillations.  Salinity effects, on the other hand, act to
increase the magnitude of the oscillations and thus act as positive
feedbacks.

A comment regarding the suitability of the mixed boundary conditions
ocean-only models for simulating the thermohaline circulation is
appropriate here.  Willebrand (1993) and Marotzke (1994) identify
certain feedbacks between the oceanic and atmospheric heat and
moisture budgets important for establishing the large-scale stability
of the underlying mean oceanic state.  Leaving out some of these
feedbacks, as done when conventional mixed boundary conditions are
employed in ocean-only models, may lead to an unrealistic stability
analysis.  Indeed, mixed boundary condition ocean-only GCMs are
perhaps more unstable than models more completely incorporating these
feedbacks (e.g., Zhang et al.\ 1993, T94, and Rahmstorf and Willebrand
1994).  In the present study, we have therefore placed the mixed
boundary condition box model in a stable regime.  Note that the
stability of the box model solution, reflected in the decay time of
the damped oscillations, is sensitive to boundary conditions through
both the temperature restoring time $\gamma_{T}^{-1}$ and salinity
forcing amplitude determined by $\gamma_{S}^{-1}$.  This result has
been seen in other studies of the THC (e.g., Zhang et al.\ 1993, Power
and Kleeman 1994, Marotzke 1994, Weaver et al.\ 1991, and T94).  In
order to study the mechanism suggested here in an ocean-only GCM, the
GCM would need to be also placed in a stable regime under mixed
boundary conditions by modifying both the salinity forcing (Weaver et
al.\ 1991, and T94) and temperature restoring time (Zhang et al.\ 1993
and Power and Kleeman 1994).  Furthermore, the GCM should be placed
near enough to the stability transition point (T94) to allow for the
non-trivial contributions of both temperature and salinity as
described here.

\section{Conclusions}
\label{section:conclusions}

We have considered the interdecadal variability in an ocean-only
four-box model of the meridional thermohaline circulation (THC) and
compared these oscillations with those found in the coupled
ocean-atmosphere model of D93.  The box model contains only advective
processes which in turn render its governing equations
(\ref{eq:T1})--(\ref{eq:S4}) nonlinear.  However, the small amplitude
($5 \% - 10 \%$ the mean circulation) response of the box model, of
interest in the present study, can be given a straightforward linear
interpretation.

The box model was placed in a stable thermally dominant mean
circulation steady state under mixed boundary conditions.  The model
linearized about this state contains a single exponentially damped
oscillatory eigenmode; all other modes are decaying and
non-oscillatory and of little importance for understanding the model's
low frequency, small amplitude oscillatory variability.  The presence
of a moderate amount of stochastic surface heating (standard deviation
$< 10\%$ the steady state heat flux), emulating the random forcing due
to atmospheric weather, drives an effectively linear box model
variability dominated by the damped oscillatory mode.  Three main
factors proved essential for the box model thermohaline oscillations:
a positive feedback from salinity produces the initial growth in the
absolute value of the anomalous fields and circulation; a negative
feedback from temperature, following the effects of salinity, is
responsible for the variability reaching a local extremum and
thereafter approaching a zero anomaly; and a time delay between the
temperature and salinity in the southern and northern boxes allowing
for the zero crossing of the oscillating anomalies.  For the stable
state considered in this paper, the temperature feedbacks dominate the
salinity feedbacks over the course of an oscillation which therefore
yield a damped oscillatory response.

Although the box model results reported here are taken from the
nonlinear model represented by the governing equations
(\ref{eq:T1})--(\ref{eq:S4}), the model's variability is of a linear
noise driven character arising from the random excitation of a single
damped oscillatory thermohaline eigenmode.  This result should be
contrasted to those ocean-only models investigating the internal
variability of the ocean which tend to focus on regimes controlled by
self-sustaining nonlinear mechanisms which do not require external
forcing (e.g., Weaver et al.\ 1991, 1993, 1994, Winton et al.\ 1993,
Moore and Reason 1993, Marotzke 1989, 1990, Yin and Sarachik 1993, Yin
1993, and Greatbatch and Zhang 1993, Cai et al.\ 1994).

The qualitative agreement of the interdecadal THC variability seen in
the stochastically driven box model and the coupled ocean-atmosphere
GCM supports the view that the coupled model's variability can be
interpreted in a linear manner analogous to that of the box model.
More precisely, the comparison supports the hypothesis that the
fundamental meridional mode of low frequency variability in the
coupled model's THC is that of a damped oscillatory thermohaline mode
driven by atmospheric weather.  It is this stable linear hypothesis
for the coupled model's variability, and its support from the box
model study, that is the main result of the present study.
Nevertheless, at the conclusion of this study, the hypothesis for the
coupled model remains speculative.  Further work is clearly necessary
to examine the linear noise-driven interpretation of both the coupled
GCM and the actual climate system.

\vskip 1.0truecm

{\bf Acknowledgments.} It is a pleasure to thank Kirk Bryan and Tom
Delworth for their valuable insights and comments given throughout
this project.  Furthermore, additional thanks go to Tom Delworth for
generously providing us with the results of D93.  Thanks go to Isaac
Held and Cecile Penland for useful comments and suggestions.  Richard
Greatbatch provided a very thorough and useful review.  Funding for
SMG is provided by a fellowship from the NOAA Postdoctoral Program in
Climate and Global Change and NOAA's Geophysical Fluid Dynamics
Laboratory.  We thank GFDL for its hospitality and support.

\newpage

\section{Bibliography}
\begin{description}

\item  Bryan, F., 1986:
High-latitude salinity effects and interhemispheric thermohaline
circulations.  {\em Nature}, {\bf 323}, 301--304.

\item Bryan, K. and F.C. Hansen, 1992: A stochastic model of North
Atlantic climate variability on a decade to century time-scale.  {\em
Proceedings of the Workshop on Decade-to-Century Time Scales of
Climate Variability}, National Research Council, Board on Atmospheric
Sciences and Climate, National Academy of Sciences, Irvine,CA,
September, 1992.

\item Cai, W., R.J. Greatbatch, and S. Zhang, Interdecadal variability
in an ocean model driven by a small zonal redistribution of the
surface buoyancy flux.  {\em Journal of Physical Oceanography} in
press.

\item Delworth, T., S. Manabe, R.J. Stouffer, 1993: Interdecadal
variations of the thermohaline circulation in a coupled
ocean-atmosphere model.  {\em Journal of Climate}, {\bf 12}, 1993--2011.

\item Greatbatch, R.J. and S. Zhang, 1994: An interdecadal oscillation
in an idealized ocean basin forced by constant heat flux.
{\em J. of Climate} in press.

\item Huang, R.X., J.R. Luyten, and H.M. Stommel, 1992: Multiple
equilibria states in combined thermal and saline circulation.  {\em
J. Phys. Oceanogr.}, {\bf 22}, 231--246.

\item Hughes, T.M.C.  and A.J. Weaver, 1994: Multiple equilibria of an
asymmetric two-basin ocean model.  {\em Journal of Physical
Oceanography}, {\bf 24}, 619--637.

\item Kloeden, and Platen, 1992: {\em Numerical Solution of Stochastic
Differential Equations}, Springer-Verlag.

\item Marotzke, J., 1989: Instabilities and multiple steady states of
the thermohaline circulation.  {\em Oceanic Circulation Models:
Combining Data and Dynamics}, D.L.T. Anderson and J. Willebrand, Eds.,
NATO ASI series, Kluwer, 501--511.

\item Marotzke, J., 1990: Instabilities and multiple equilibria of the
thermohaline circulation.  Ph.D. thesis. Berlin Instit Meereskunde
Kiel, 126 pp.

\item Marotzke, J., P. Welander, and J. Willebrand, 1988:
Instabilities and multiple steady states in a meridional-plane model
of the thermohaline circulation.  {\em Tellus}, {\bf 40A}, 162--172.

\item Marotzke, J., 1994: Ocean models in climate problems.  {\em
Ocean Processes in Climate Dynamics: Global and Mediterranean
Examples}, P. Malanotte-Rizzoli and A.R. Robinson, eds., Kluwer
Academic Publishers.

\item Mikolajewicz, U., and E. Maier-Reimer, 1990: Internal secular
variability in an ocean general circulation model.  {\em Climate
Dynamics}, {\bf 4}, 145--156.

\item Moore, A.M. and C.J.C. Reason, 1993: The response of a global
general circulation model to climatological surface boundary
conditions for temperature and salinity.  {\em J. Phys. Oceanogr.},
{\bf 23}, 300--328.

\item Mysak, L.A., T.F. Stocker, and F. Huang, 1993:
Century-scale variability in a randomly forced, two-dimensional
thermohaline ocean circulation model.
{\em Climate Dynamics}, {\bf 8}, 103--116.

\item Penland, C., 1989: Random forcing and forecasting using
principal oscillation pattern analysis. {\em Monthly Weather Review},
{\bf 117}, 2165--2185.

\item Power, S.B. and R. Kleeman, 1993: Multiple equilibria in a
global ocean general circulation model.  {\em J. Phys. Oceanogr.},
{\bf 23}, 1670--1681.

\item Rahmstorf, S. and J. Willebrand, 1994: The role of temperature
feedback in stabilising the thermohaline circulation.  {\em Journal of
Physical Oceanography} in press.

\item Ruddick, B.\ and L. Zhang, 1994: On the qualitative behaviour an
non-oscillation of Stommel's thermohaline box model.  {\em Journal of
Climate} submitted.

\item Stommel, H., 1961: Thermohaline convection with two stable
regimes of flow.  {\em Tellus}, {\bf 13}, 224--230.

\item Tziperman, E. and K. Bryan, 1993: Estimating global air-sea
fluxes from surface properties and from climatological flux data using
an oceanic general circulation model. {\em Journal of Geophysical
Research}, {\bf 98}, 22629--22644.

\item Tziperman, E., R. Toggweiler, Y. Feliks, and K. Bryan, 1994:
Instability of the thermohaline circulation with respect to mixed
boundary conditions: Is it really a problem for realistic models?
{\em J. Phys. Oceanogr.}, {\bf 24}, 217--232.


\item Walin, G., 1985: The thermohaline circulation and the control of
ice ages.  {\em Paleogeography, Paleoclimatology, and Paleoecology},
{\bf 50}, 323--332.

\item Weaver, A.J. and E.S. Sarachik 1991: The role of mixed boundary
conditions in numerical models of the ocean's climate.  {\em
J. Phys. Oceanogr.}, {\bf 21}, 1470--1493.

\item Weaver, A.J., E.S. Sarachik, and J. Marotzke, 1991: Freshwater
flux forcing of decadal and interdecadal oceanic variability. {\em
Nature}, {\bf 353}, 836--838.

\item Weaver, A.J. and T.M.C. Hughes, 1992: Stability and variability
of the thermohaline circulation and its link to climate. {\em Trends
in Phys. Oceanography}, {\bf 1}, 15--70.

\item Weaver, A.J., J. Marotzke, P.F. Cummins, and E.S. Sarachik,
1993a: Stability and variability of the thermohaline circulation.
{\em J. of Phys. Ocean.}, {\bf 23}, 39--60.

\item Weaver, A.J., S.M. Aura, and P.G. Myres, 1994: Interdecadal
variability in an idealized model of the North Atlantic.  {\em Journal
of Geophysical Research},  {\bf 89} 12423--12441.

\item Weisse, R., U. Mikolajewicz, and E. Maier-Reimer, 1994: Decadal
variability of the North Atlantic in an ocean general circulation
model.  {\em Journal of Geophysical Research}, {\bf 89} 12411--12421.

\item Welander, P., 1986: Thermohaline effects in the ocean
circulation and related simple models.  {\em Large-Scale Transport
Processes in the Oceans and Atmosphere}, D.L.T. Anderson and J.
Willebrand, Eds., NATO ASI series, Reidel.

\item Willebrand, J, 1993: Forcing the ocean with heat and freshwater
fluxes. In {\em Energy and Water Cycles in the Climate System},
E. Raschke, ed., Springer-Verlag.

\item Winton, M. and E.S. Sarachik, 1993: Thermohaline oscillations
induced by strong steady salinity forcing of ocean general circulation
models.  {\em J. Phys. Oceanogr.}, {\bf 23}, 1389--1410.

\item Wright, D. G. and T.F. Stocker, 1991:
A zonally averaged ocean model for the thermohaline circulation. Part
I: model development and flow dynamics.  {\em Journal of Physical
Oceanography}, {\bf 21}, 1713--1724.

\item  Yin, F.L., 1993: A mechanistic model of ocean interdecadal
thermohaline oscillations.  JISAO preprint 233, June 1993.

\item Yin, F.L., and E.S. Sarachik, 1993: On interdecadal oscillations
in a sector ocean general circulation model: advective and convective
processes.  {\em J. Phys. Oceanogr.} in press.

\item Zhang, S., R.J. Greatbatch, and C.A. Lin, 1993: A reexamination
of the polar halocline catastrophe and implications for coupled
ocean-atmosphere modeling.  {\em J. Phys. Oceanogr.}, {\bf 23},
287--299.

\end{description}



\newpage

\end{document}